\documentclass[twocolumn]{aastex62}
\usepackage{array,multirow}
\usepackage{ragged2e}
\usepackage{float}
\usepackage{lmodern}
\usepackage{mwe}
\usepackage{graphicx}
\usepackage{comment}

\newcommand{\um}{$\mu$m}
\newcommand{\SE}{SESNe}

\newcommand{\lam}{$\lambda$}
\newcommand{\kms}{km~s$^{-1}$}

\shorttitle{NIR Spectroscopy of SESNe}
\shortauthors{Shahbandeh et al.}

\begin{document}

\title{Carnegie Supernova Project-II: Near-infrared Spectroscopy of Stripped-Envelope Core-Collapse Supernovae\footnote{This paper includes data gathered with the 6.5 meter \textit{Magellan} Telescopes at Las Campanas Observatory, Chile.}}

\correspondingauthor{M. Shahbandeh}
\email{melissa.shahbandeh@gmail.com}

\author[0000-0002-9301-5302]{M. Shahbandeh}
\affil{Department of Physics, Florida State University, 77 Chieftan Way, Tallahassee, FL 32306, USA}

\author[0000-0003-1039-2928]{E. Y. Hsiao}
\affil{Department of Physics, Florida State University, 77 Chieftan Way, Tallahassee, FL 32306, USA}

\author[0000-0002-5221-7557]{C. Ashall}
\affil{Institute for Astronomy, University of Hawai'i at Manoa, 2680 Woodlawn Dr., Hawai'i, HI 96822, USA}

\author[0000-0001-8290-2881]{J. Teffs}
\affil{Astrophysics Research Institute, Liverpool John Moores University, IC2, Liverpool Science Park, 146 Brownlow Hill, Liverpool L3 5RF, UK}

\author[0000-0002-4338-6586]{P. Hoeflich}
\affil{Department of Physics, Florida State University, 77 Chieftan Way, Tallahassee, FL 32306, USA}

\author[0000-0003-2535-3091]{N. Morrell}
\affil{Carnegie Observatories, Las Campanas Observatory, Casilla 601, La Serena, Chile}

\author[0000-0003-2734-0796]{M. M. Phillips}
\affil{Carnegie Observatories, Las Campanas Observatory, Casilla 601, La Serena, Chile}

%alphabetical

\author[0000-0003-0227-3451]{J. P. Anderson}
\affil{European Southern Observatory, Alonso de C\'ordova 3107, Casilla 19, Santiago, Chile}

\author[0000-0001-5393-1608]{E. Baron}
\affil{Homer L. Dodge Department of Physics and Astronomy, 440 West Brooks Street, Room 100, Norman, OK 73019, USA}

\author[0000-0003-4625-6629]{C. R. Burns}
\affil{Observatories of the Carnegie Institution for Science, 813 Santa Barbara St, Pasadena, CA 91101, USA}

\author[0000-0001-6293-9062]{C. Contreras}
\affil{Carnegie Observatories, Las Campanas Observatory, Casilla 601, La Serena, Chile}

\author[0000-0002-2806-5821]{S. Davis}
\affil{Department of Physics, University of California, 1 Shields Avenue, Davis, CA 95616-5270, USA}

\author[0000-0002-0805-1908]{T. R. Diamond}
\affil{Private Astronomer}

\author[0000-0001-5247-1486]{G. Folatelli} 
\affil{Facultad de Ciencias Astron\'{o}micas y Geof\'{i}sicas, Universidad Nacional de La Plata, Instituto de Astrof\'{i}sica de La Plata (IALP), CONICET, Paseo del Bosque S/N, B1900FWA La Plata, Argentina}

\author[0000-0002-1296-6887]{L. Galbany}
\affil{Institute of Space Sciences (ICE, CSIC), Campus UAB, Carrer de Can Magrans, s/n, E-08193 Barcelona, Spain}

\author[0000-0002-8526-3963]{C. Gall}
\affil{DARK, Niels Bohr Institute, University of Copenhagen, Jagtvej 128, 2200 Copenhagen, Denmark}

\author[0000-0001-8341-1478]{S. Hachinger}
\affil{Leibniz Supercomputing Centre (LRZ), Boltzmannstr. 1, D-85748 Garching, Germany}

\author[0000-0002-3415-322X]{S. Holmbo}
\affil{Department of Physics and Astronomy, Aarhus University, Ny Munkegade, DK-8000 Aarhus C, Denmark}

\author[0000-0001-6209-838X]{E. Karamehmetoglu}
\affil{Department of Physics and Astronomy, Aarhus University, Ny Munkegade, DK-8000 Aarhus C, Denmark}

\author[0000-0002-5619-4938]{M. M. Kasliwal}
\affil{Division of Physics, Mathematics, and Astronomy, California Institute of Technology, Pasadena, CA 91125, USA}

\author[0000-0002-1966-3942]{R. P. Kirshner}
\affil{Harvard-Smithsonian Center for Astrophysics, 60 Garden Street, Cambridge, MA 02138, USA}
\affil{Gordon and Betty Moore Foundation, 1661 Page Mill Road, Palo Alto, CA 94304, USA}

\author[0000-0002-6650-694X]{K. Krisciunas}
\affil{George P. and Cynthia Woods Mitchell Institute for Fundamental Physics and Astronomy, Department of Physics and Astronomy, Texas A\&M University, College Station, TX, 77843, USA}

\author[0000-0001-8367-7591]{S. Kumar}
\affil{Department of Physics, Florida State University, 77 Chieftan Way, Tallahassee, FL 32306, USA}

\author[0000-0002-3900-1452]{J. Lu}
\affil{Department of Physics, Florida State University, 77 Chieftan Way, Tallahassee, FL 32306, USA}

\author{G. H. Marion}
\affil{Department of Astronomy, University of Texas, 1 University Station C1400, Austin, TX 78712, USA}

\author[0000-0001-6876-8284]{P. A. Mazzali}
\affil{Astrophysics Research Institute, Liverpool John Moores University, IC2, Liverpool Science Park, 146 Brownlow Hill, Liverpool L3 5RF, UK}
\affil{Max-Planck Institut fur Astrophysik, Karl-Schwarzschild-Str. 1, D-85741 Garching, Germany}

\author[0000-0001-6806-0673]{A. L. Piro}
\affil{Observatories of the Carnegie Institution for Science, 813 Santa Barbara St, Pasadena, CA 91101, USA}

\author[0000-0003-4102-380X]{D. J. Sand}
\affiliation{Department of Astronomy and Steward Observatory, University of Arizona, 933 N Cherry Avenue, Tucson, AZ 85719, USA}

\author[0000-0002-5571-1833]{M. D. Stritzinger}
\affil{Department of Physics and Astronomy, Aarhus University, Ny Munkegade, DK-8000 Aarhus C, Denmark}

\author[0000-0002-8102-181X]{N. B. Suntzeff}
\affil{George P. and Cynthia Woods Mitchell Institute for Fundamental Physics and Astronomy, Department of Physics and Astronomy, Texas A\&M University, College Station, TX, 77843, USA}

\author[0000-0002-2387-6801]{F. Taddia}
\affil{Department of Physics and Astronomy, Aarhus University, Ny Munkegade, DK-8000 Aarhus C, Denmark}

\author[0000-0002-9413-4186]{S. A. Uddin}
\affil{George P. and Cynthia Woods Mitchell Institute for Fundamental Physics and Astronomy, Department of Physics and Astronomy, Texas A\&M University, College Station, TX, 77843, USA}

\begin{abstract}
We present 75 near-infrared (NIR; 0.8$-$2.5~\um) spectra of 34 stripped-envelope core-collapse supernovae (SESNe) obtained by the Carnegie Supernova Project-II (CSP-II), encompassing optical spectroscopic Types IIb, Ib, Ic, and Ic-BL.
The spectra range in phase from pre-maximum to 80~days past maximum.
This unique data set constitutes the largest NIR spectroscopic sample of SESNe to date.
NIR spectroscopy provides observables with additional information that is not available in the optical. 
Specifically, the NIR contains the resonance lines of \ion{He}{1} and allows a more detailed look at whether Type Ic supernovae are completely stripped of their outer He layer. 
The NIR spectra of SESNe have broad similarities, but closer examination through statistical means reveals a strong dichotomy between NIR ``He-rich'' and ``He-poor'' SNe.
These NIR subgroups correspond almost perfectly to the optical IIb/Ib and Ic/Ic-BL types, respectively.
The largest difference between the two groups is observed in the 2~\um\ region, near the \ion{He}{1}~\lam2.0581~\um\ line. 
The division between the two groups is \emph{not} an arbitrary one along a continuous sequence.
Early spectra of He-rich SESNe show much stronger \ion{He}{1}~\lam2.0581~\um\ absorption compared to the He-poor group, but with a wide range of profile shapes. 
The same line also provides evidence for trace amounts of He in half of our SNe in the He-poor group.
\end{abstract}


\section{Introduction} \label{sec:intro}

Stripped-envelope core-collapse supernovae (SESNe) mark the death of massive stars whose progenitors have been stripped of their outer hydrogen and sometimes even helium envelopes \citep{Clocchiatti_1996}.
These SESNe are classified into three categories: Types IIb, Ib, Ic \citep{Filippenko_1995,Matheson_2001}. 
Type IIb supernovae (SNe~IIb) show weak H lines at early times that disappear later and stronger He lines at late times, SNe~Ib show no H but prominent He, and SNe~Ic show neither H nor He.
Broad-line SNe~Ic (SNe~Ic-BL) constitute a unique subgroup in Type Ic that show broad spectral features and are often associated with long gamma-ray bursts \citep[e.g.][]{Iwamoto_1998, Ashall_2019}.
These classifications are made mainly using optical spectral features.

It is not yet clear if \SE\ arise from single evolved massive stars or interacting binary systems or both.
It is also unknown whether these SN subclasses represent distinct explosion mechanisms or progenitor systems.
For He-rich SNe~Ib and IIb, there have been several detections of the progenitor stars, all of which were determined to have $M_{\rm ZAMS}<$ 25~M$_{\sun}$ \citep[e.g.][]{Cao_2013,Fremling_2014,Bersten_2014, Prentice_2018}.
For SN~Ib~iPTF13bvn, a possible binary companion was detected. 
However, iPTF13bvn is a peculiar Type Ib \citep{Kuncarayakti_2015,Folatelli_2016}. 
Progenitor detections for SNe~IIb are more common than SNe~Ib, including SN~1993J: \citep{Aldering_1994, Cohen_1995}, SN~2008ax: \citep{Li_2008}, SN~2013df: \citep{VanDyk_2014}, and SN~2011dh: \citep{Maund_2011, VanDyk_2011, Bersten_2012, Ergon_2015}.

To date, there has yet to be a conclusive direct detection of the progenitor of a SN~Ic, although \citet{VanDyk_2018} and \citet{Rho_2021} found possible evidence for the progenitors of SNe~Ic SN~2017ein and SN~2020bvc, respectively. 
A study by \citet{Beasor_2021} suggested that for masses below 27~M$_{\sun}$, the single star pathway is unlikely for the production of Wolf-Rayet stars and SESNe. 
Commonly, WO and WC Wolf-Rayet stars have been proposed as possible progenitors for SNe~Ic \citep[e.g.,][]{Crowther_2007, Dessart_2011, Yoon_2015}.

It has long been debated as to how \SE\ lose their outer envelopes, especially since their progenitor type and configuration are uncertain.
Some of the scenarios that could cause copious mass loss include: i) eruptions  \citep{Smith_2006b, Smith_2006a}, ii) radiation driven winds  \citep{Heger_2003b, Smith_2006b, Pauldrach_2012}, iii) stripping from common envelope in a binary system  \citep{Nomoto_1984, Nomoto_1987, Podsiadlowski_1993, Woosley_1995, Wellstein_1999, Wellstein_2001, Pols_2002, Heger_2003b, Podsiadlowski_2004}, iv) fast rotations in Be stars \citep{Massa_1975, Kogure_1982, Slettebak_1988, Meynet_2005a, Meynet_2005b, Owocki_2006}, or a combination of the above. 
Each of these scenarios has a different dependence on and relation with the initial mass of the progenitor. 

It is not clear if there is a continuum of properties between sub-types of \SE, or if each sub-type points to a separate physical origin.
The continuity between Type Ib and Ic hints at possible connections.
The residual \ion{He}{1} in SNe~Ic has been studied for some time but with uncertain results
\citep[e.g.][]{Hachinger_2012, Piro_2014, Prentice_2018, Dessart_2020, Williamson_2021}.

\cite{Teffs_2020_a} concluded that one cannot hide a significant mass of He without line formation at some epoch, and the low He mass in bare C/O core models at 8 $\times\ 10^{51}$ ergs (8 foe) still produce NIR \ion{He}{1} lines but not strong optical \ion{He}{1} lines. 
As the mass of He in the He-poor/He-rich models increases, the \ion{He}{1} lines become more apparent.
Challenges in identifying residual He are exacerbated by the fact that
the \ion{Na}{1}~D lines are in the same wavelength region ($\sim$ 5875$-$5890\AA) as the strongest \ion{He}{1} line in the optical.
In most SNe~Ic, a small amount of He may be present but the lack of excitation by gamma-rays could cause some of the He to be effectively ``hidden'', \citep{Lucy_1991,Hachinger_2012,Piro_2014}.
With NIR spectroscopy, we can begin to assess the nature of SESNe, for example, the \ion{He}{1} lines. 
The NIR region contains two strong \ion{He}{1} lines: \lam1.0830~\um\ and \lam2.0581~\um. 
The current published sample size of NIR spectra is limited, and there have been only a few objects with rapid-cadence coverage \citep[e.g.][]{Taubenberger_2011, Drout_2016}.
Here, we present and analyze 75 NIR spectra of 34 \SE, with the aim of providing a better understanding of the origins of \SE. 
The spectra were obtained as part of the second phase of {\it The Carnegie Supernova Project}  \citep[CSP-II;][]{Phillips_2019}.

The main focus of this work is to identify ions responsible for the prominent features near 1~\um\ and 2~\um, specifically presence of He. In Section~\ref{sec:sample}, the observations and the sample characteristics are presented.
Section~\ref{sec:lineid} describes how the dominant ions for the most prominent spectral features are identified.
Section~\ref{sec:measurements} presents our spectroscopic measurements, including velocities and pseudo-equivalent widths for several lines. 
Principal component analysis (PCA) of the data is presented in Section~\ref{sec:PCA}.
The limitations of our interpretations of the data and comparisons to models are discussed in Section~\ref{sec:discussion}. 
Finally, a summary of conclusions is given in Section~\ref{sec:summary}.

%#################################################
\section{Observations and Sample} \label{sec:sample}

%CSP follow up
We present the largest homogeneous NIR spectroscopic sample of \SE\ published to date obtained as part of CSP-II, a four-year NSF-supported project between 2011 and 2015.
The project covered four observing campaigns spanning 7-8 months each from October to May centered on the Chilean summer.
The NIR spectroscopy program was a vital component of the CSP-II \citep{Hsiao_2019} with the vast majority of the entire CSP-II data obtained using the Folded-port Infrared Echellette \citep[FIRE;][]{Simcoe_2013} mounted at Baade telescope at Las Campanas Observatory (LCO). 
All 75 NIR spectra of this sample were taken with FIRE.
The data can be downloaded from our website.\footnote{https://csp.obs.carnegiescience.edu/data}
The spectra were observed, reduced, and corrected for telluric absorption following the steps described by \citet{Hsiao_2019}.

\begin{figure*}
    \epsscale{1.2}
    \centering
    \plotone{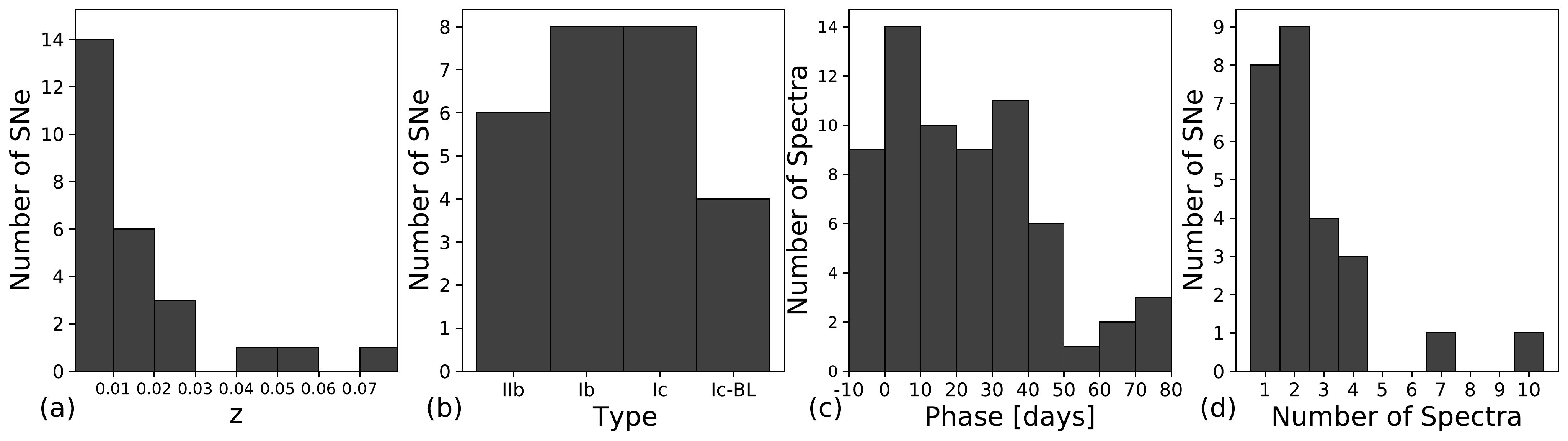}
    \caption{Sample characteristics expressed in four histograms: a) the distribution of the sample SNe with respect to heliocentric redshift, the SNe with a redshift higher than 0.04 are SN~2013co, LSQ13ddu, and LSQ13lo,  b) the distribution of SNe with respect to Type, c) the distribution of the NIR spectra with respect to phase, and d) the number of SESNe with and without time-series observations. The majority of the SNe only have one NIR spectrum. The SNe with the best coverage are SN~2013ge with 10 and SN~2014L with 7 NIR spectra.\label{fig:SNvsSpec}}
\end{figure*}

%redshifts
The main aim of the NIR spectroscopy program was to obtain high signal-to-noise ratio (S/N) spectra of nearby SNe in order to construct a SN~Ia spectral template for K-correction estimates as well as to study the physical origins of SNe of all types.
Therefore, the majority of the CSP-II targets are nearby and at lower redshifts.
The mean heliocentric redshift of our SESNe sample is $z=0.015$ with the distribution shown in Figure~\ref{fig:SNvsSpec} (panel a).
The redshifts of the host galaxies were mainly obtained from the NASA/IPAC Extragalatic Database (NED).\footnote{http://ned.ipac.caltech.edu/} 
Two exceptions are OGLE-2014-SN-067 and SN~2013fq.
The host galaxy redshift of OGLE-2014-SN-067 was determined using narrow host [\ion{O}{3}] and [\ion{N}{2}] emission lines present in an optical spectrum taken on 2014 August 29 UT \citep{ATEL6430} with the Wide Field Spectrograph (WiFeS) on the Australian National University (ANU) 2.3-m telescope \citep{Childress_2016}.
The WiFeS spectrum of SN~2013fq taken on 2013 September 19 UT also showed narrow emission lines from the host galaxy indicating a heliocentric redshift of 0.0113 \citep{CBET3665}.
Table~\ref{tab:redshift_JD} lists host names and heliocentric redshifts.

%classification
The classifications for the SNe in our sample were determined using optical spectra as close to maximum light as possible.
Note that these classification determinations were independent of any NIR spectroscopic information.
The details of the optical classifications and the full list of the spectra used are presented in Appendix~\ref{appendix:Classification}.
Each of the types IIb, Ib, Ic, and Ic-BL is represented in the sample with a handful of SNe (Figure~\ref{fig:SNvsSpec}, panel b).

%phase
In this work, the phases of the NIR spectra are labeled relative to the time of $B$-band maximum unless otherwise noted.
For 11 of the objects in the sample, optical photometric follow-up observations, obtained with the 1-m Swope Telescope at LCO, were available to determine the time of maximum. 
We further relied on previously published light curves for 10 other SNe.
The $B$-band light curves (unless otherwise noted) were then interpolated with Gaussian Processes using the package included in SuperNovae in object oriented Python\footnote{https://csp.obs.carnegiescience.edu/data/snpy} \citep[{\tt SNooPy};][]{Burns_2011} in order to determine the time of maximum.
For five SNe in the sample with no usable photometry, we used their optical spectra and the SuperNova IDentification code {\tt SNID v5.0} \citep{Blondin&Tonry_2007} with Ib/c templates from \cite{Liu_Modjaz_2014} to estimate the phase. 
The SNID-determined phases were assigned an uncertainty of $2.0$~days, approximately equal to the largest uncertainty in the time of maximum determined using light curves. 
Panel c of Figure~\ref{fig:SNvsSpec} shows the distribution of spectra with respect to phase. 
The time of maximum and the method for determining it are listed in Table~\ref{tab:redshift_JD} for each SN. 
Note that this table includes 26 SNe with both optical classification and a time of maximum estimate. 
The eight SNe in our sample without either optical classification or a time of maximum are listed in Appendix~\ref{appendix:exclude} and were excluded from our analyses.
The journal of observations for the 67 NIR spectra of 26 SNe included in our analyses is shown in Table~\ref{tab:journal}.

\begin{deluxetable*}{cccccchh}
\tablecaption{Host-galaxy and SN phase information \label{tab:redshift_JD}}
\tablehead{
\colhead{SN} &\colhead{Host} & \colhead{Redshift\tablenotemark{a}}& \colhead{No. of Spec} & \colhead{MJD$_{B_{\rm max}}$} &\colhead{Source of MJD$_{B_{\rm max}}$\tablenotemark{b}}\\
\colhead{} &\colhead{} & \colhead{}& \colhead{} & \colhead{day} &\colhead{}}
\startdata
ASASSN-14az 			&   WISEA J234448.30-020653.1       &	0.0067&  	3 	&  56815.0$\pm$2.0   	&SNID&   23:44:48.00 & 	$-$02:07:03.17\\
LSQ13abf				&	WISEA J114906.60+191006.2	    &	0.0208&	1	&  56417.3$\pm$0.5 &\cite{Stritzinger_2020}
&	11:49:06.62	& +19:10:10.71\\
LSQ13cum				&	WISEA J001507.27-240543.4		&	0.0260&	3	& 56604.7$\pm$0.5 &CSP-II&	00:15:07.51	&	$-$24:05:38.42\\
LSQ13ddu				&	WISEA J035849.18-292508.5		&	0.0584 &	2	& 56626.0$\pm$0.2$^{c}$&\cite{Clark_2020}&	03:58:49.11	&	$-$29:25:11.93\\
LSQ13doo	            &	UGCA 247						&	0.0065& 	2 	& 56599.0$\pm$2.0 &SNID&	11:48:45.78	&	$-$28:17:31.20\\
LSQ13lo					&	WISEA J142638.77+033018.1		&	0.0717&	1	  & 56377.9$\pm$0.2 &CSP-II&	14:26:38.78	&	+03:30:11.60\\
LSQ14akx				&	WISEA J143405.47-073936.5		&	0.0228&	4	& 56750.5$\pm$0.3 &CSP-II&	14:34:05.75	&	$-$07:39:36.50\\
iPTF13bvn				&	NGC 5806						&	0.0045&	3	& 56473.5$\pm$0.5 &\cite{Fremling_2016}&	15:00:00.18	&	+01:52:53.50\\
SN 2011hs				&	IC 5267							&	0.0057&	2	&55884.8$\pm$0.1 &CSP-II&	14:56:23.64	&	$-$37:36:18.00\\
SN 2012P				&	NGC 5806						&	0.0045&	2	& 55948.5$\pm$0.5  &\cite{Fremling_2016}&	06:42:42.55	&	$-$27:26:49.80\\
SN 2012ap				&	NGC 1729						&	0.0121&	1	& 55973.5$\pm$0.2 &CSP-II&	14:59:59.04	&	+01:53:24.40\\
SN 2012au				&	NGC 4790						&	0.0045&	2	& 55885.0$\pm$1.0  &\cite{Bufano_2014}&	22:57:11.77	&	$-$43:23:04.80\\
SN 2012hf				&	NGC 3469						&	0.0155&	2	& 56259.1$\pm$1.4&CSP-II&	12:54:52.18	&	$-$10:14:50.20\\
SN 2012hn				&	NGC 2272						&	0.0071&1& 56032.5$\pm$1.0&\cite{Valenti_2014}&	10:13:47.95	&	+03:26:02.60\\
SN 2013ak				&	ESO 430- G 020					&	0.0034&	4	& 56366.1$\pm$1.8 &\cite{Pessi_2019}&	08:07:06.69	& $-$28:03:10.10\\
SN 2013co				&	WISEA J125550.49+303042.5  		&	0.0497&	1	& 56425.0$\pm$2.0 &SNID&	12:55:50.51	&	+30:30:41.50\\
SN 2013dk				&	NGC 4038						&	0.0055&	3	& 56476.0$\pm$2.0$^{d}$&\cite{Elias-Rosa_2013}&	12:01:52.72	&	$-$18:52:18.30\\
SN 2013ek				&	NGC 6984						&	0.0156&	2	&56496.9$\pm$1.1&CSP-II&	20:57:53.92	&$-$51:52:24.80\\
SN 2013fq				&	-								&	0.0113&	1	&56550.0$\pm$2.0	& SNID&	19:59:07.95	&	$-$55:55:46.60\\
SN 2013ge				&	NGC 3287						&	0.0043&	10	& 56615.7$\pm$1.5&\cite{Drout_2016}&	10:34:48.46	&	+21:39:41.90\\
SN 2014L				&	M99								&	0.0080&	7	& 56693.2$\pm$0.2&\cite{Zhang_2018}&	12:18:48.68	&	+14:24:43.50\\
SN 2014ad				&	MRK 1309						&	0.0057&	4	& 56735.1$\pm$0.2&\cite{Sahu_2018}&	11:57:44.44	&	$-$10:10:15.70\\
SN 2014ar				&	ESO 266- G 015					&	0.0106&	2	& 56770.8$\pm$0.1&CSP-II&	11:40:58.90	&	$-$44:29:05.70\\
SN 2014df				&	NGC 1448						&	0.0039&2	&56799.0$\pm$2.0	& SNID&	03:44:23.99	&	$-$44:40:08.10\\
SN 2014eh	            &	NGC 6907						&	0.0106&	1 	& 56973.2$\pm$0.2&CSP-II&	20:25:03.86	&	$-$24:49:13.30\\
SN 2015Y	            &	NGC 2735						&	0.0082& 	1 	& 57136.3$\pm$0.2&CSP-II&	09:02:37.87	&	+25:56:04.20\\
\enddata
\tablenotetext{a}{Heliocentric redshift.}
\tablenotetext{b}{Source of the light curves used to determine the time of maximum if available; or using SNID in case of light curve unavailability.}
\tablenotetext{c}{With respect to LSQ-$gr$ band maximum.}
\tablenotetext{d}{With respect to $V$-band maximum.}
\end{deluxetable*}

\begin{figure*}
    \epsscale{1.}
    \centering
    \plotone{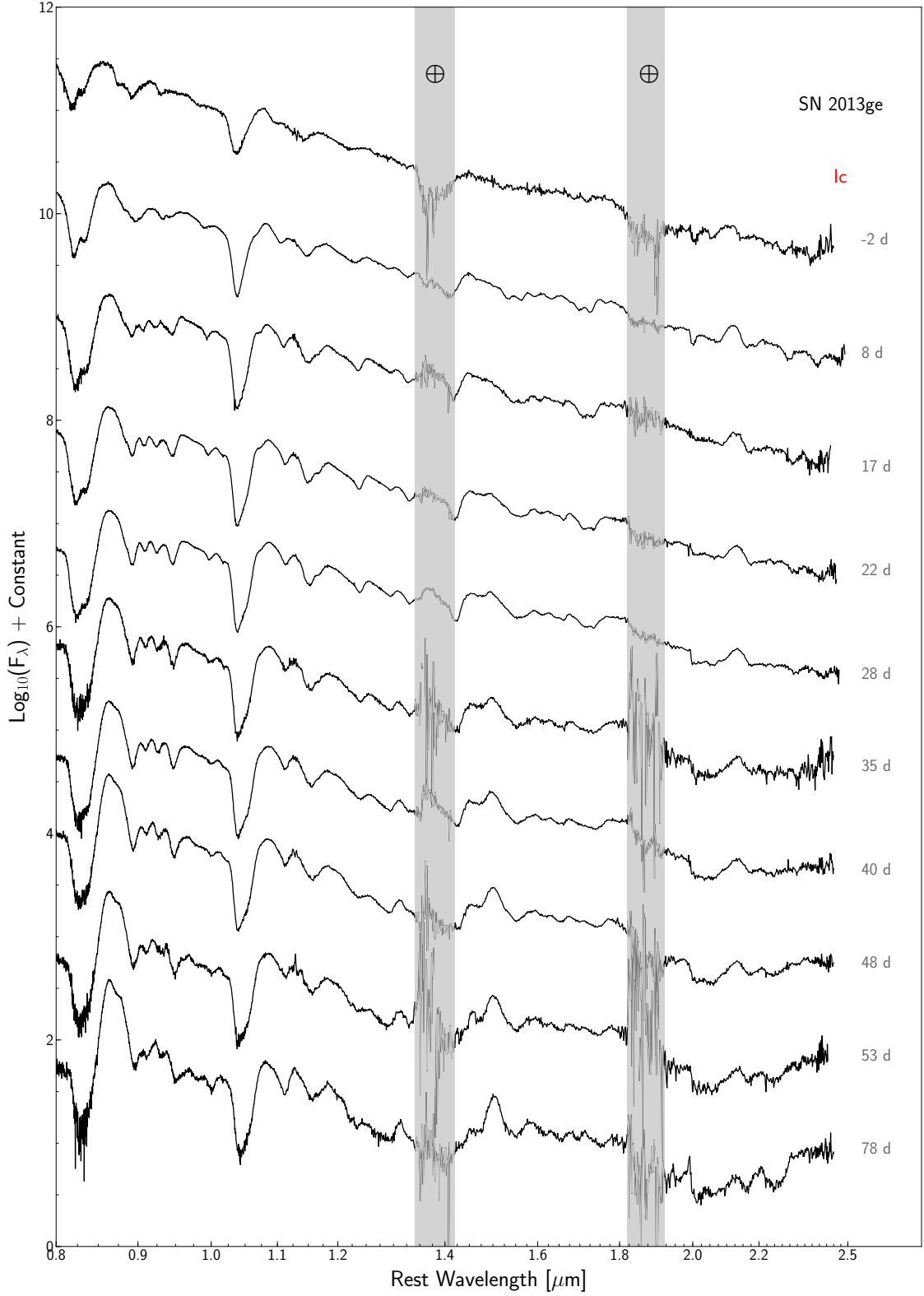}
    \caption{Time-series NIR spectra of the Type Ic SN~2013ge. This SN has the best time coverage of the sample. The two grey bands mark regions that have high telluric absorptions from the atmosphere. The number on the right side of the each spectrum is the phase relative to $B$-band maximum. \label{fig:13ge}}
\end{figure*}

\begin{figure*}
    \epsscale{1.}
    \centering
    \plotone{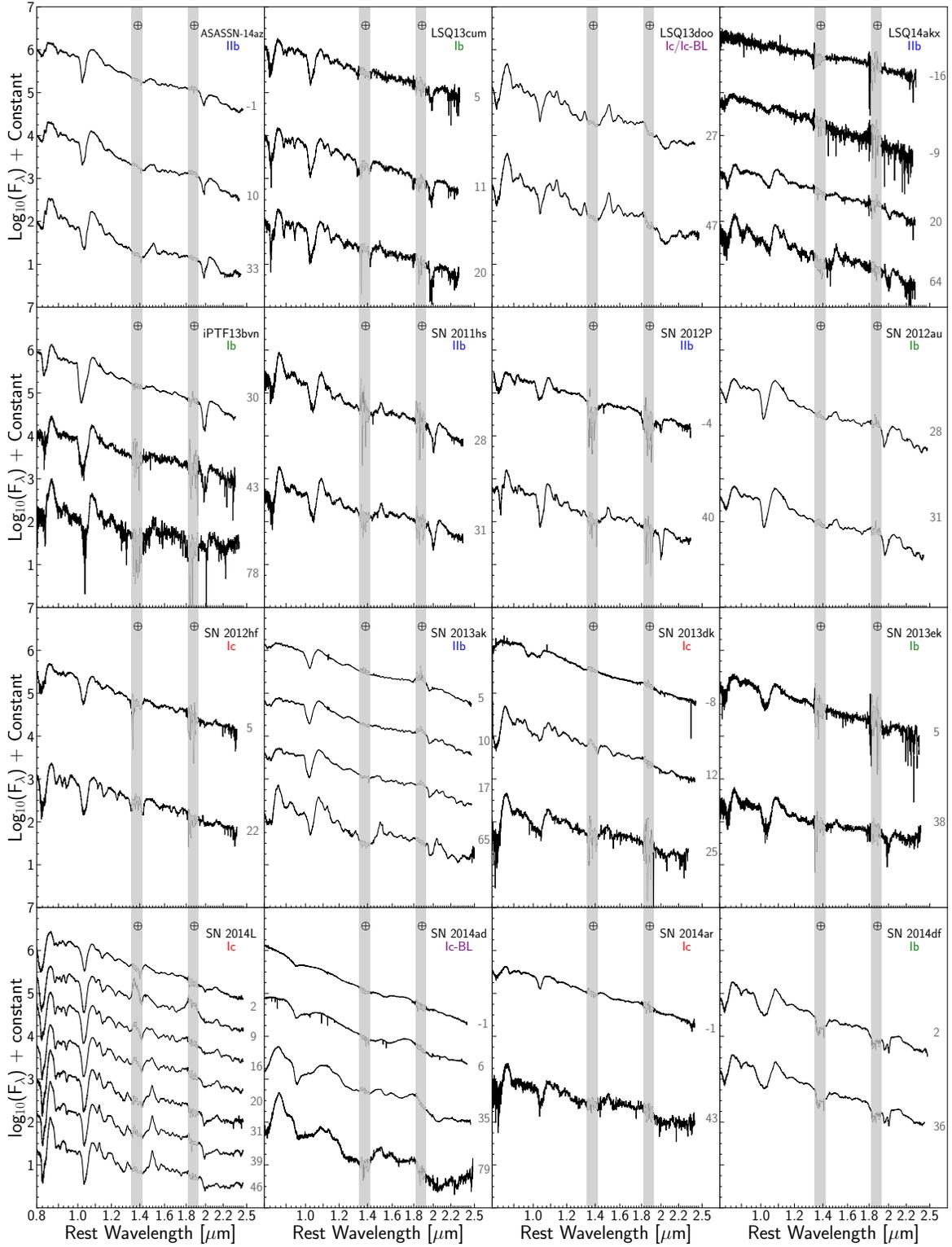}
    \caption{Time-series NIR spectra of the SNe with two or more spectra. Note that the time series of SN~2013ge is shown separately in Figure~\ref{fig:13ge}. The phase relative to maximum light of each spectrum and the type of each SN are also labeled. The wavelength regions with strong telluric absorptions are marked with vertical grey bands.
    \label{fig:paperplots_4X4}}
\end{figure*}

\begin{figure}
    \epsscale{1.22}
    \centering
    \plotone{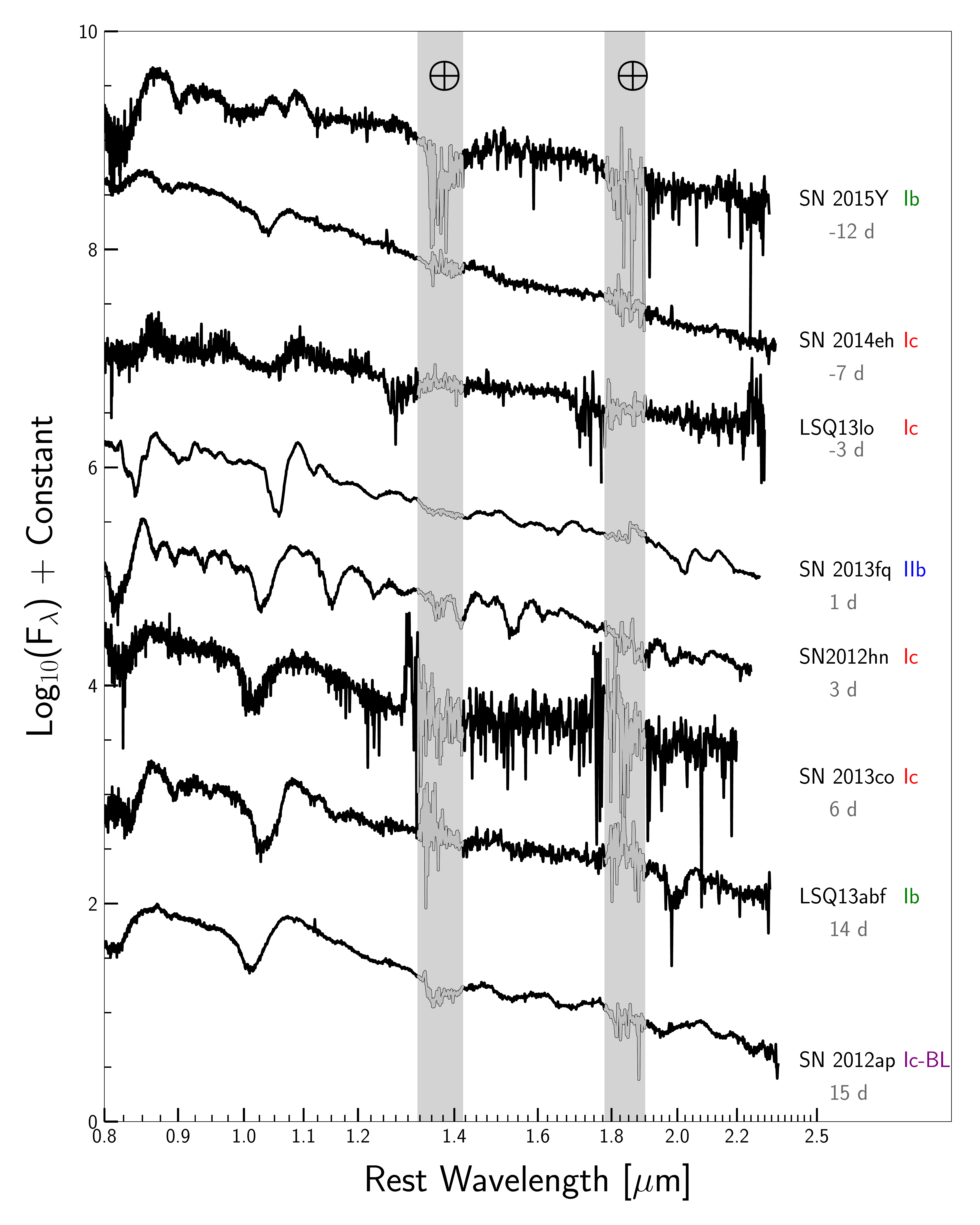}
    \caption{Spectra of the SNe with a single spectrum within the sample that have optical spectra and/or photometry available for phase estimation and spectroscopic classification.
    The feature blueward of the telluric region for SN 2013co is likely an artifact of the observation and not real.\label{fig:all_spec-bad}}
\end{figure}

%time series
This sample includes time-series NIR spectroscopic sequences for 9 SNe with 3 NIR spectra or more (panel d of Figure~\ref{fig:SNvsSpec}).
SN~2013ge, a SN~Ic, has the best time coverage in the sample with 10 spectra spanning from near maximum light to approximately 80 days past maximum (Figure~\ref{fig:13ge}). 
Time-series spectroscopy for all remaining SESNe with two or more spectra are presented in Figure~\ref{fig:paperplots_4X4}.
There are several SNe with only one NIR spectrum each. These spectra were mostly taken for classification purposes and are presented in Figure~\ref{fig:all_spec-bad}. 

%blind vs targeted
Striving for an unbiased sample of SNe Ia, CSP-II prioritized the follow-up of objects discovered by untargeted transient surveys \citep{Phillips_2019}. 
Of the 26 SNe we used in our analyses, 46\% were discovered by untargeted surveys, including the All-Sky Automated Survey for SuperNovae \citep[ASAS-SN;][]{Shappee_2014, Holoien_2017}, Catalina Real Time Transit Survey \citep[CRTS;][]{Djorgovski_2011}, La Silla-Quest Low Redshift Survey \citep[LSQ;][]{Baltay_2013}, and the Palomar Transient Factory \citep[PTF;][]{Law_2009}.

\startlongtable
\begin{deluxetable*}{ccccccchh}
\tabletypesize{\scriptsize}
\tablecaption{Journal of observations\label{tab:journal}}
\tablehead{
\colhead{SN} & \colhead{UT date} & \colhead{MJD} & \colhead{Phase} & 
\colhead{Telluric STD} & \colhead{Airmass}& \colhead{Total exp time}\\
\colhead{} & \colhead{} & \colhead{day} & \colhead{day} & 
\colhead{} & \colhead{}& \colhead{s}
}
\startdata
ASASSN-14az 			& 2014-06-06 & 56814.9 & $-$1.0	 	%& 
& HD4329 		& 1.4 & 507.2 & Baade+FIRE \\
								ASASSN-14az		& 2014-06-17 & 56824.9 & +10.4 		 
										& HIP116886 	& 1.1 & 507.2 & Baade+FIRE \\
								ASASSN-14az		& 2014-07-10 & 56847.9 & +33.4 	 
										& HD8325 		& 1.1 & 1014.4 & Baade+FIRE \\
LSQ13abf 				& 2013-05-19 & 56430.6 & +14.2 	 
& HIP59861 	& 1.6 & 1268.0 & Baade+FIRE \\
LSQ13cum 				& 2013-11-14 & 56609.7 & +5.5 		 
& HD4329 		& 1.5 & 1268.0 & Baade+FIRE \\
								LSQ13cum		& 2013-11-20 & 56615.7 & +11.4 	 
										& HD8325 		& 1.4 & 1268.0 & Baade+FIRE \\
								LSQ13cum		& 2013-11-30 & 56625.6 & +20.0 	 
										& HD2339 		& 1.2 & 1014.4 & Baade+FIRE \\
LSQ13ddu 				& 2013-12-09 & 56635.2 & +9.0 	 
& HD27803 	& 1.4 & 1902.0 & Baade+FIRE \\
							LSQ13ddu			& 2013-12-14 & 56640.2 & +14.0 	 
										& HD33243 	& 1.3 & 1521.6 & Baade+FIRE\\
LSQ13doo 	& 2013-11-30 & 56625.8 & +27.3 	 
& HD106010	& 1.5 & 507.2 & Baade+FIRE \\
                                  LSQ13doo          & 2013-12-20 & 56645.8 & +47.3 	 
& HD96781 	& 1.4 & 317.0 & Baade+FIRE \\
LSQ13lo  				& 2013-03-25 & 56375.8 & $-$2.6 		 
& HD117248 	& 1.2 & 1394.8 & Baade+FIRE \\
LSQ14akx 				& 2014-03-18 & 56733.8 & $-$16.2	 
& HD157334 	& 1.1 & 1014.4 & Baade+FIRE \\
								LSQ14akx		& 2014-03-25 & 56740.8 & $-$9.1 	 
										& HD129684 	& 1.1 & 2028.8 & Baade+FIRE \\
                            	LSQ14akx		& 2014-04-23 & 56769.7 & +20.0 	 
                            			& HD125062 	& 1.1 & 1521.6 & Baade+FIRE \\
                            	LSQ14akx		& 2014-06-06 & 56813.6 & +63.6 	 
                            			& HD125062	& 1.2 & 760.8 & Baade+FIRE \\
iPTF13bvn				& 2013-07-16 & 56488.6 & +30.2 	 
& HD134240 	& 1.6 & 951.0 & Baade+FIRE & TJD\\
                           		iPTF13bvn		& 2013-07-29 & 56501.5 & +43.1 	 
                           				& HD100852 	& 1.3 & 1268.0 & Baade+FIRE & EH\\
                           		iPTF13bvn		& 2013-09-02 & 56536.5 & +78.1 	 
                           				& HD125062 	& 1.8 & 760.8 & Baade+FIRE & MK\\
SN 2011hs				
										& 2011-12-18 & 55912.6 & +28.5 	 
										& HD221793 	& 1.4 & 465.2 & Baade+FIRE & EH\\
                            	SN 2011hs		& 2011-12-21 & 55916.0 & +31.0	 
                            			& HD223296 	& 1.7 & 507.2 	& Baade+FIRE & EH\\
SN 2012P  				& 2012-01-19 & 55944.9 & $-$3.6 	 
& HD23722 	& 1.7 & 380.4  & Baade+FIRE & MK, EH\\
SN 2012P		& 2012-03-03 & 55988.8 & +40.3 	 
										& HD147295 	& 1.2 & 1268.0 & Baade+FIRE & EH\\
SN 2012ap 				& 2012-03-03 & 55988.5 & +14.8 	 
& HD290322 	& 1.3 & 1268.0 & Baade+FIRE & EH\\
SN 2012au 				& 2012-04-18 & 56034.7 & +28.2 	 
& HD109587 	& 1.2 & 960.0 & Baade+FIRE & RF\\
								SN 2012au 		& 2012-04-21 & 56037.7 & +31.2 	 
										& HD109587 	& 1.3 & 960.0 & Baade+FIRE & RF\\
SN 2012hf 				& 2012-12-03 & 56263.8 & +5.2 		 
& HD90738 	& 1.2 & 2219.0 & Baade+FIRE & EH\\
								SN 2012hf 		& 2012-12-20 & 56280.7 & +22.2 	 
										& HD105992 	& 1.2 & 3170.0 & Baade+FIRE & MP, NM\\
SN 2012hn 				& 2012-04-19 & 56036.0 & +3.0 		 
& HD50963 	& 1.2 & 1200.0 & Baade+FIRE & RF\\
SN 2013ak 				& 2013-03-20 & 56370.6 & +4.9 	 
& HD67526 	& 1.0 & 380.4 	& Baade+FIRE & EH\\
								SN 2013ak		& 2013-03-25 & 56375.5 & +10.0 	 
										& HD77562 	& 1.0 & 380.4 	& Baade+FIRE & EH\\
					       		SN 2013ak		& 2013-04-01 & 56382.6 & +17.0 	 
					       				& HD67638 	& 1.5 & 253.6 	& Baade+FIRE & EH\\
                            	SN 2013ak		& 2013-05-18 & 56430.4 & +64.8 	
                            			& HD66018 	& 1.1 & 507.2 & Baade+FIRE & HM, EH\\
SN 2013co 				& 2013-05-19 & 56430.6 & +6.1 	 
& HIP59861 	& 2.0 & 1268.0 & Baade+FIRE & HM, EH\\
SN 2013dk 				& 2013-06-25 & 56467.5 & $-$8.5 	
& HD94359 	& 1.1 & 360.0  & Baade+FIRE & MR\\
					        	SN 2013dk		& 2013-07-15 & 56488.4 & +12.5 	 
					        			& HD97919 	& 1.1 & 1014.4 & Baade+FIRE & TJD\\
                            	SN 2013dk		& 2013-07-29 & 56501.5 & +25.5 	 
                            			& HD100852 	& 1.5 & 1268.0 & Baade+FIRE & EH\\
SN 2013ek				& 2013-07-29 & 56501.7 & +4.8 		
& HD200523 	& 1.1 & 1902.0 & Baade+FIRE & EH\\
								SN 2013ek		& 2013-09-02 & 56536.7 & +38.0 	 
										& HD203999 	& 1.2 & 507.2 & Baade+FIRE & MK\\
SN 2013fq				& 2013-09-18 & 56553.0 & +1.0 		 
& HD196106 	& 1.1 & 507.2 & Baade+FIRE & YB\\
SN 2013ge 			& 2013-11-20 & 56615.9 & $-$1.9 	 
& HD284582 	& 2.1 & 507.2 & Baade+FIRE & EH, TD\\
								SN 2013ge		& 2013-11-30 & 56625.9 & +7.8 		 
										& HD106010 	& 1.9 & 380.4 & Baade+FIRE & TD, EH\\
					        	SN 2013ge		& 2013-12-09 & 56634.8 & +16.7 	 
					        			& HD96781 	& 1.9 & 380.4 & Baade+FIRE & TD, EH\\
                              	SN 2013ge		& 2013-12-14 & 56639.8 & +21.7 	
                              			& HD96781 	& 1.8 & 507.2 & Baade+FIRE & EH\\
                            	SN 2013ge		& 2013-12-20 & 56645.8 & +27.7 	 
                            			& HD96781 	& 1.8 & 507.2 & Baade+FIRE & EH\\
                            	SN 2013ge		& 2013-12-27 & 56652.8 & +34.7 	 
                            			& HD96781 	& 1.9 & 507.2 & Baade+FIRE & EH\\
                            	SN 2013ge		& 2014-01-01 & 56657.8 & +39.6 	 
                            			& HD96781 	& 1.6 & 507.2 & Baade+FIRE & EH, MK\\
                            	SN 2013ge		& 2014-01-09 & 56665.9 & +47.7 	 
                            			& HD96781 	& 1.6 & 380.4 & Baade+FIRE & EH\\
                            	SN 2013ge		& 2014-01-14 & 56670.8 & +52.6	 
                            			& HD96781 	& 1.6 & 507.2 & Baade+FIRE & EH\\
                            	SN 2013ge		& 2014-02-08 & 56695.8 & +77.8 	 
                            			& HD96781 	& 1.7 & 380.4 & Baade+FIRE & HM\\
SN 2014L 				& 2014-02-08 & 56695.9 & +1.6 		 
& HD96781 	& 1.5 & 380.4 & Baade+FIRE & HM\\
								SN 2014L		& 2014-02-15 & 56702.8 & +8.6 	 
										& HD96781 	& 1.4 & 570.6 & Baade+FIRE & HM, SH\\
					        	SN 2014L		& 2014-02-22 & 56709.8 & +15.5 	
					        			& HD96781 	& 1.4 & 380.4 & Baade+FIRE & EH\\
                            	SN 2014L		& 2014-02-27 & 56714.8 & +20.5  
                            			& HIP59861 	& 1.4 & 507.2 & Baade+FIRE & EH\\
                            	SN 2014L		& 2014-03-10 & 56725.7 & +31.4 	
                            			& HD96781 	& 1.5 & 507.2 & Baade+FIRE & MP, EH\\
					        	SN 2014L		& 2014-03-18 & 56733.7 & +39.4 	  
					        			& HD96781 	& 1.4 & 507.2 & Baade+FIRE & EH\\
                            	SN 2014L		& 2014-03-25 & 56740.7 & +46.4 	 
                            			& HD96781 	& 1.4 & 507.2 & Baade+FIRE & NM, EH\\
SN 2014ad 				& 2014-03-18 & 56733.6 & $-$1.0 	 
& HD105992 	& 1.3 & 507.2 & Baade+FIRE & EH\\
								SN 2014ad		& 2014-03-25 & 56740.7 & +6.2 	& HD97919 	& 1.2 & 507.2 & Baade+FIRE & NM, EH\\
					        	SN 2014ad		& 2014-04-23 & 56769.7 & +35.0 	 
					        			& HD97919 	& 1.1 & 1014.4 & Baade+FIRE & NM\\
                            	SN 2014ad		& 2014-06-06 & 56813.6 & +79.0 	 
                            			& HD122435 	& 1.6 & 760.8 & Baade+FIRE & HM\\
SN 2014ar 				& 2014-04-23 & 56769.7 & $-$0.6 	 
& HD105011 	& 1.2 & 1521.6 & Baade+FIRE & NM\\
								SN 2014ar		& 2014-06-06 & 56813.6 & +43.3 	 
										& HD122435 	& 1.6 & 760.8 & Baade+FIRE & HM\\
SN 2014df 				& 2014-06-06 & 56814.4 & +2.0 		 
& HD18620 	& 2.5 & 507.2 & Baade+FIRE & HM\\
								SN 2014df		& 2014-07-10 & 56848.4 & +36.0 	 
										& HD23722 	& 1.3 & 1014.4 & Baade+FIRE & NM\\
SN 2014eh 	& 2014-11-05 & 56965.5 & $-$7.2 	 
& HD202633 	& 1.1 & 1014.4 & Baade+FIRE \\
SN 2015Y 	& 2015-04-12 & 57123.5 & $-$12.2 	 
& HD96781 	& 1.7 & 1521.6 & Baade+FIRE \\
\enddata
\end{deluxetable*}

%%%%%%%%%%%%%%%%%%%%%%%%%%%%%%%%%%%%%%%%%%%%%%%%%%%%%%%%%%%%%%%%

%%%%%%%%%%%%%%%%%%%%%%%%%%%%%%%%%%%%
\section{Line Identification}\label{sec:lineid}

In this section, we attempt to identify the ions responsible for the main features in the NIR spectra and describe their evolution over time. This can help us investigate properties of the progenitor stars of SESNe. 
The line identifications given in \cite{Taubenberger_2006, Mazzali_2010, Taubenberger_2011, Bufano_2014} and \cite{Ergon_2015} were used as a guide.
The most prominent P Cygni line of SESNe is just redward of 1~\um, regardless of the SN class.
To discern whether the ion responsible for this 1~\um\ feature is \ion{He}{1} \lam1.0830~\um, we would expect to see other \ion{He}{1} lines, therefore the absorption near 2~\um\ becomes especially important.
We will refer to these two as the 1~\um\ and 2~\um\ features, respectively, in the remaining text.
The line identifications are presented in Figure~\ref{fig:lineid}.

\subsection{Pa$\gamma$ \lam1.0938~\um, Pa$\beta$ \lam1.2818~\um, Pa$\alpha$ \lam1.8750~\um, and Br$\gamma$ \lam2.1655~\um}
The most prominent features observed in SNe~II are produced by hydrogen. 
%Pa$\gamma$
Pa$\gamma$ \lam1.094~\um\ is located in the 1~\um\ region that is heavily dominated by the \ion{He}{1} \lam1.0830~\um\ line, thus not easily discernible. 
%Pa$\beta$
Pa$\beta$ is detected in the $<$~29~days spectra of all SNe~IIb in the sample, except ASASSN-14az, a SN that seems to be featureless around 1.2~\um.
In the case of SN~2013ak, assuming this line is Pa$\beta$, the emission feature seems to be blueshifted in the day~5 spectrum (shown in the left panel of Figure~\ref{fig:lineid}) but not in the day~17 one (shown in the right panel of Figure~\ref{fig:lineid}).
Also, noting that a P Cygni profile is detected at the same wavelength ($\sim$1.26~\um) in the majority of the SNe~Ib and Ic in the sample as well.
In SNe~IIb, this feature is concurrent with Pa$\alpha$, H$\alpha$, H$\beta$, and H$\gamma$. 
Thus, we identify it as Pa$\beta$.
Conversely, in SNe~Ib and Ic, this feature appears with \ion{C}{1} lines in both the optical and NIR. 
Additionally, we do not detect any other H lines in these SNe. Therefore, we identify this feature as \ion{C}{1} \lam1.2614~\um.
%Pa$\alpha$
Pa$\alpha$ \lam1.8750~\um\ lies in a region with heavy telluric absorption, even when there is adequate signal from the SN after telluric correction, Pa$\alpha$ \lam1.8750~\um\ is not easily detectable. 
However, we detect Pa$\alpha$ in SN~2013ak and LSQ14akx.
%Br$\gamma$
The Br$\gamma$ \lam2.1655~\um\ line is located near where we see the emission feature of the \ion{He}{1} \lam2.0581~\um\ line, therefore it may be heavily dominated by \ion{He}{1}.
Due to the presence of \ion{He}{1} in SNe~IIb and Ib, it is difficult to distinguish between a SN~IIb and a SN~Ib using NIR spectroscopy alone. 
In Figure~\ref{fig:lineid}, we have shown a Type II SN with \ion{H}{1} lines marked in blue for comparison with SESNe.

\subsection{\ion{He}{1} \lam1.0830~\um, \lam1.7002~\um, and \lam2.0581~\um}
\ion{He}{1} \lam1.0830~\um\ forms a distinct and prominent P Cygni in the spectra of SNe~IIb and Ib, and possibly contributes to the 1~\um\ feature in SNe~Ic. 
This profile is produced by the 2$^{3}$S-2$^{3}$P transition.
\ion{He}{1} \lam1.0830~\um\ line is about $5-15$ times stronger than the \ion{He}{1} \lam2.0581~\um\ line and much stronger than any optical \ion{He}{1} lines.
The \ion{He}{1} \lam2.0581~\um\ feature can also be easily detected in the spectra of SNe~IIb and Ib. 
This line is formed by 2$^{1}$S-2$^{1}$P transition. 
The other \ion{He}{1} at \lam1.7002~\um\ is much weaker than the other mentioned \ion{He}{1} lines, but it can be found in some of the SNe in the sample. 
\ion{He}{1} lines are identified in orange in Figure~\ref{fig:lineid}.
One of the main goals of this work is to determine whether He is present in SNe~Ic.
This topic is explored in detail in Section~\ref{sec:discussion}.

\subsection{\ion{C}{1} \lam0.9093~\um, \lam0.9406~\um, \lam0.9658~\um, \lam1.0693~\um, \lam1.2614~\um, and \lam2.1259~\um}
One or more of the three \ion{C}{1} lines: \lam0.9093~\um, \lam0.9406~\um, \lam0.9658~\um\ are detected in most spectra of this sample regardless of the type, even though they are much weaker than the \ion{C}{1} \lam1.0693~\um\ line.
They may be blended with \ion{O}{1} \lam0.9264~\um\ or \ion{Mg}{2} \lam0.9227~\um, making it challenging to distinguish their contribution to the weak lines in this region. 
On the other hand, the \ion{C}{1} \lam1.0693~\um\ line is one of the strongest lines in the NIR. 
It also coincides with the strong 1~\um\ feature present in all SESNe in the sample.
We therefore consider \ion{C}{1} \lam1.0693~\um\ one of the possible main contributors to the 1~\um\ feature.
A detailed examination of this profile and contributions from \ion{He}{1} \lam1.0830~\um\ and \ion{C}{1} \lam1.0693~\um\ are presented in Section~\ref{sec:discussion}.
We detect a feature in the majority of the SNe in this sample around 1.26~\um, that could be formed by \ion{C}{1} \lam1.2614~\um.
Other \ion{C}{1} lines may also form weaker features, such as \lam1.1330~\um, \lam1.1754~\um, and \lam1.4543~\um.
The \ion{C}{1} \lam2.1259~\um\ line may contribute to the absorption feature near 2~\um, also discussed in Section~\ref{sec:discussion}. In Figure~\ref{fig:lineid}, \ion{C}{1} lines are marked in purple.

\subsection{\ion{O}{1} \lam0.9264~\um\ and \lam1.1290~\um}
The \ion{O}{1} \lam1.1290~\um\ line is the strongest \ion{O}{1} line in the NIR.
It can be identified in most spectra as a weak absorption, situated just redward of the prominent 1~\um\ feature.
Over time, this feature becomes more apparent.
The \lam0.9264~\um\ feature is another \ion{O}{1} line that is present in many of the SNe in the sample.
The \ion{O}{1} lines are marked in brown in Figure~\ref{fig:lineid}.

\subsection{\ion{Mg}{1} \lam1.1828~\um, \lam1.4878~\um, and \lam1.5033~\um}
\ion{Mg}{1} \lam1.4878~\um\ and \lam1.5033~\um\ form a narrow emission-like feature that is characteristic of most evolved NIR spectra of SESNe.
They sometimes form a single emission and sometimes form a ``double-horned'' emission with peaks that coincide with the rest wavelengths of the two lines. 
The absorption component of the P Cygni profile may be blended with the emission component of \ion{C}{1} \lam1.454~\um, be affected by the telluric region or, it could be a result of de-excitation from higher levels. 
The feature emerges around 30 days past maximum for nearly all SNe in the sample and becomes stronger over time. 
The weaker \ion{Mg}{1} \lam1.1828~\um\ line lies in the telluric region, and it is therefore not easily detectable.
The \ion{Mg}{1} lines are marked in magenta in Figure~\ref{fig:lineid}.

\subsection{\ion{Mg}{2} \lam0.9227~\um, \lam1.0092~\um, \lam1.0927~\um, and \lam2.1369~\um}
\ion{Mg}{2} \lam1.0927~\um\ may be another major contributor to the 1~\um\ feature. 
Its effect is examined in detail in Sections~\ref{sec:PCA} and \ref{sec:discussion}.
The weaker \ion{Mg}{2} \lam0.9227~\um\ and \lam1.0092~\um\ may be blended with \ion{C}{1} and \ion{O}{1} lines as mentioned in previous subsections.
\ion{Mg}{2} \lam2.1369~\um\ may also be a contributor to the 2~\um\ feature.
Its separation from \ion{He}{1} \lam2.0581~\um\ demonstrates a potential advantage of this region for identifying the ions that are present.

\subsection{\ion{S}{1} \lam0.9223~\um\ and \lam1.0457~\um}
\ion{S}{1} \lam0.9223~\um\ coincides with \ion{Mg}{2} \lam0.9227~\um\ and cannot be separated.
\ion{S}{1} \lam1.0457~\um\ may also contribute to the prominent 1~\um\ feature.
This line is at the bluest wavelength of all the candidates considered for the 1~\um\ feature.
 
\subsection{\ion{Ca}{2} $\lambda\lambda\lambda$0.8538, 0.8662, 0.8921~\um, \lam1.1839~\um, and \lam1.1950~\um}
\begin{figure*}
\epsscale{1}
\centering
\plotone{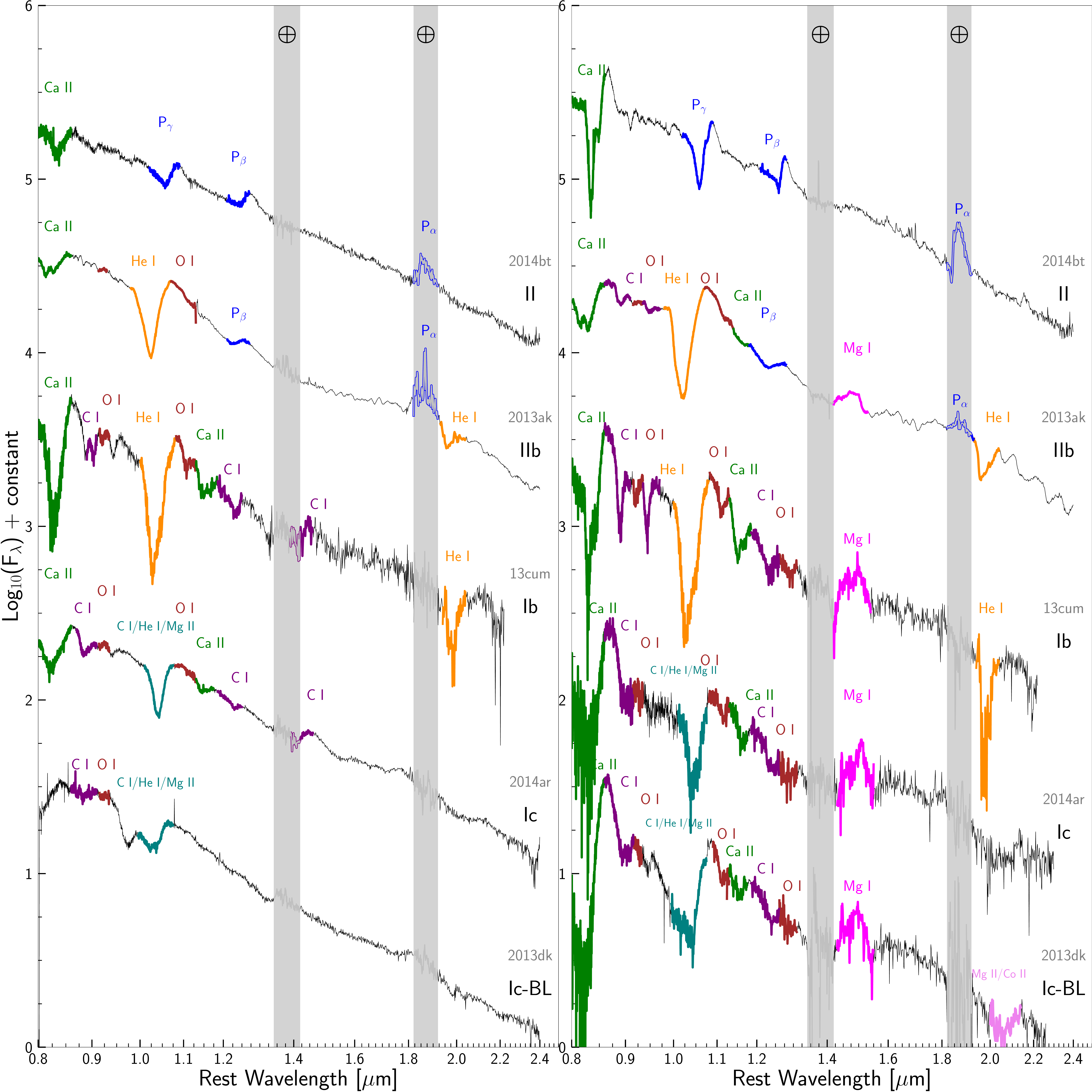}
\caption{An atlas of NIR spectra of SESNe with absorption feature of prominent lines labeled for comparison. The SNe names and types are labeled on the right hand side of each spectrum. The left and right panels show spectra around maximum light and around 30 days past maximum, respectively. The gray bands mark the regions with high telluric absorption. The Type II SN 2014bt shown at the top of the figure is for comparison.
\label{fig:lineid}}
\end{figure*}

The \ion{Ca}{2} infrared triplet is present in all of the NIR spectra in this sample. 
This feature is weak before and around maximum light and becomes progressively stronger over time.
The \ion{Ca}{2} \lam1.1839~\um\ and \lam1.1950~\um\ lines may also be present but they are not as prominent as the infrared triplet.
The \ion{Ca}{2} lines are shown in green in Figure~\ref{fig:lineid}.

%%%%%%%%%%%%%%%%%%%%%%%%
\section{Spectroscopic Measurements}
\label{sec:measurements}

%intro
To quantify the attributes of the strongest spectral features observed, we measured the pseudo-equivalent widths (pEW) and the velocities at their absorption minima. 
In particular, we focus on the prominent absorption feature near 1~\um\ and the weaker absorption feature near 2~\um\ to discern whether these can be attributed to \ion{He}{1} lines.
In this section, the details of these measurements are described.

\subsection{Equivalent width} 
\label{subsec:pEW}

%method
The pEW is used to quantify the strength of a spectral absorption feature.
We adopted the formalism for pEW outlined by \citet{Folatelli_2004} and \citet{Garavini_2007}.
As the feature boundaries are sometimes ambiguous, the boundary ranges were set manually for each spectrum.
We defined the continuum as the straight line connecting the boundaries on the blue and red sides of the absorption feature and used it to normalize the spectrum before the area is integrated. 
To estimate the uncertainty in the pEW measurement, 100 realizations of simulated spectra were created by varying the flux at each pixel randomly and independently based on its flux error.
Furthermore, the blue and red boundaries that define the continuum were also varied randomly for each realization within an average range of 0.006~\um\ around the chosen boundaries.
The median absolute deviation of the 100 pEW measurements multiplied by 1.48 was then taken as the 1-$\sigma$ error.

\begin{figure*}
    \centering
    \epsscale{1.2}
    \plotone{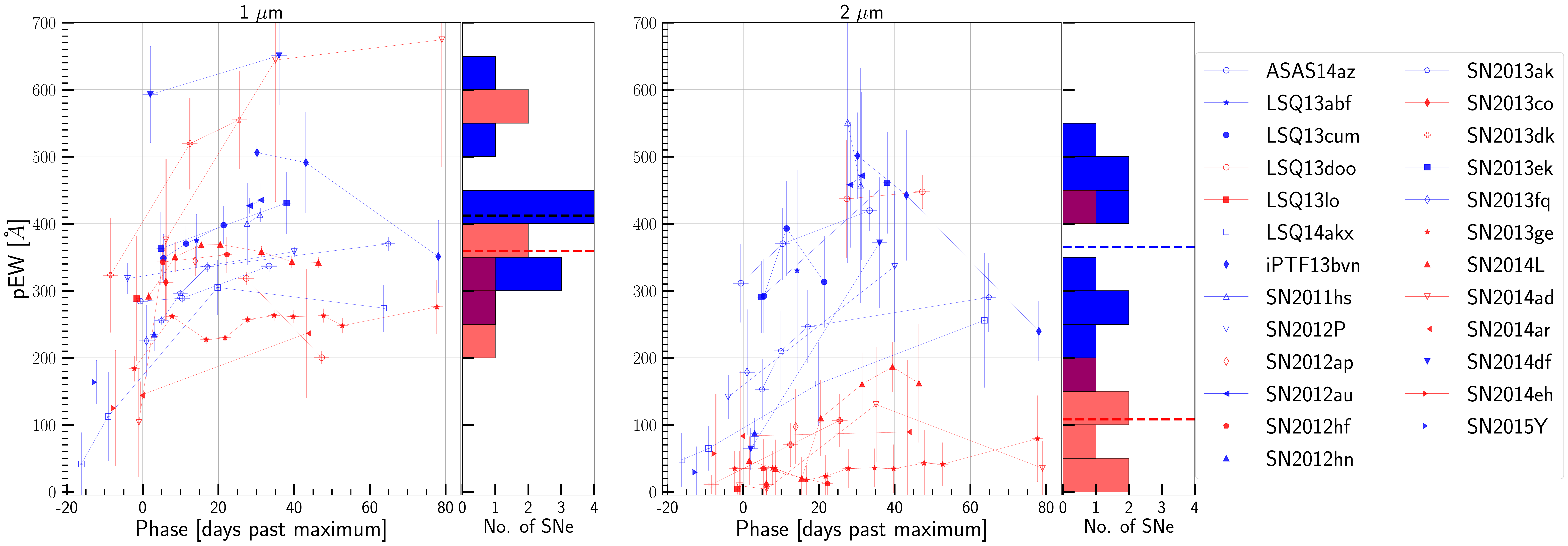}
    \caption{The pEW measurements of the absorption features near 1~\um\ (left panel) and 2~\um\ (right panel). The SN~IIb/Ib and SN~Ic/Ic-BL groups are shown in blue and red, respectively, with filled symbols representing Type Ib/Ic and open symbols representing Type IIb/Ic-BL. To compare the pEW distributions of the SN~IIb/Ib versus SN~Ic/Ic-BL groups, histograms are shown for SNe with available pEW measurements interpolated to 30~days past maximum. The blue and red dashed lines on the histograms mark the median pEW for SNe~IIb/Ib and SNe~Ic/Ic-BL at day~30, respectively.}
    \label{fig:pEW}
\end{figure*}

%result
All the spectra in our sample show strong absorption features near 1~\um\ regardless of the optical spectral type, and most show a weaker absorption near the 2~\um\ region.  
Without identifying the ions responsible for these features, we simply present the time evolution of the pEWs in Figure~\ref{fig:pEW}.  
There are large ranges of pEWs for both the 1~\um\ and the 2~\um\ features, spanning from a few tens to 500\AA. 
In SNe~IIb, both features have monotonically increasing pEWs as the spectra evolve with time.
On average, SNe~IIb and Ib show a slightly stronger 1~\um\ features than SNe~Ic and Ic-BL. 
However, if we disregard SNe~Ic-BL, the difference between the two groups becomes much more apparent. 
Also note that, the pEW measurements of the 1~\um\ feature for SN~2014L, a well-sampled SN~Ic, are situated in the average of pEWs of the SN~IIb/Ib group.

%He-rich and He-poor
There exists a much stronger dichotomy between the SN~IIb/Ib and SN~Ic/Ic-BL groups in the strength of the 2~\um\ feature compared to the 1~\um\ region.
While the pEWs of SNe~IIb/Ib span a wide range of $\sim150-500$~\AA, those of SNe~Ic/Ic-BL are largely confined to low values of $\lesssim150$~\AA\ with LSQ13doo and SN2013co being the exceptions.
The 2~\um\ feature clearly divides the sample into two distinct groups: NIR ``He-rich'' and ``He-poor'' groups with the pEW staying above and below 150$~$\AA\ over time, respectively.
The NIR He-rich and He-poor groups correspond almost perfectly with the optical SN~IIb/Ib and SN~Ic/Ic-BL groups, respectively.

\subsection{Velocity} 
\label{subsec:velocity}

%profile
In order to measure the blueshift velocity of a spectral feature, the wavelength of the absorption minimum is determined by fitting Gaussian functions to the profile via non-linear least-square minimization \citep{Newville_2014}.
Again, we focus on the absorption features near 1~\um\ and 2~\um.
In most of our spectra, these features cannot be adequately described by a single Gaussian function as indicated by the reduced $\chi^2$ which were $\gg1$ in these fits.
We found that a two-Gaussian function provides the best results for most of our spectra, although there exist a variety of profile shapes.
This may indicate the presence of multiple ions or lines formed in large optical depth for the strong 1~\um\ feature.
Detailed discussions are presented in Section~\ref{sec:discussion}.

%method
We adopted the same profile boundaries and definition of pseudo-continuum as those used for measuring the pEW.
The fit parameters are the wavelength center, width, and depth for each Gaussian component. 
The wavelength center of the component with the deepest absorption was then taken for the velocity estimate.
Each two-Gaussian fit to the 1~\um\ and 2~\um\ features is presented in Appendix~\ref{Appendix:fits}.
If a fit region has a median S/N ratio of S/N$<5$, the spectrum is excluded from the fit.
This applies to 4 spectra in our sample with low signals in the $K$ band.
The uncertainty in the wavelength of the absorption minimum and thus the velocity is determined in a similar fashion as that for pEW.
Realizations were created to account for the effects of flux errors and uncertain boundary selections.
The wavelength of the absorption minimum is determined for each of the 100 realizations.
The median absolute deviation of the absorption minimum multiplied by 1.48 was then taken as the 1-$\sigma$ uncertainty.

\begin{figure*}
    \centering
    \epsscale{1.2}
    \centering
    \plotone{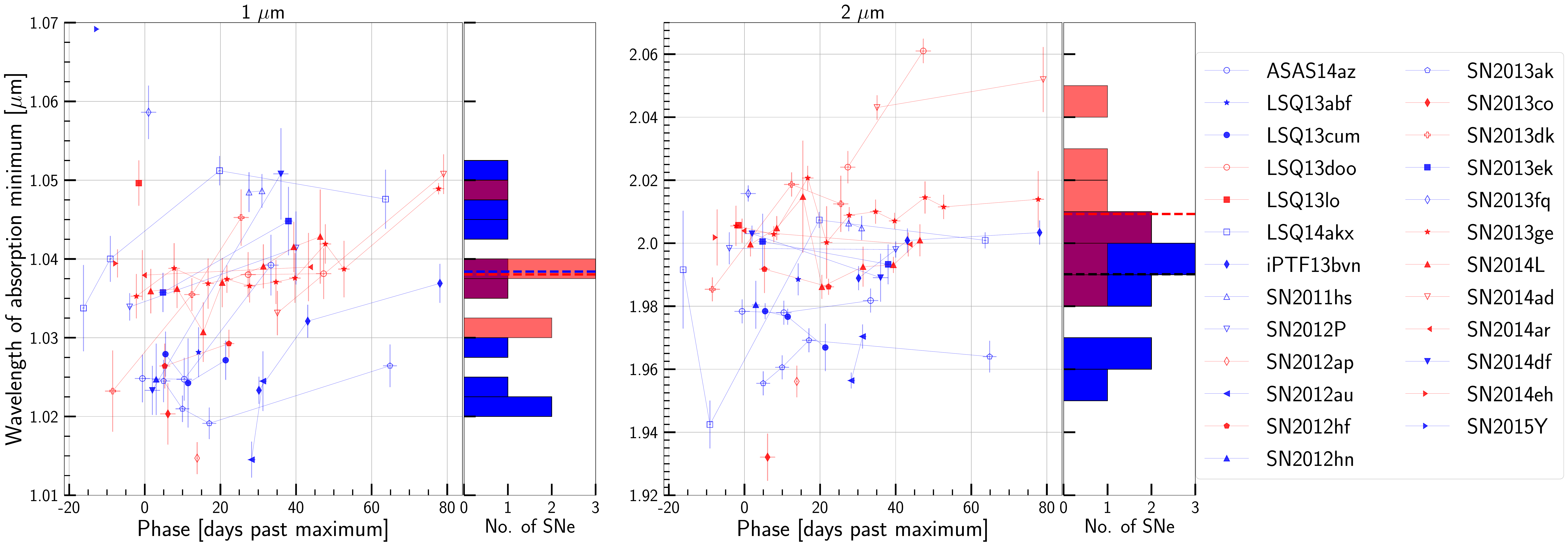}
    \caption{The rest-frame wavelengths of the absorption minima for the 1~\um\ (left panel) and 2~\um\ (right panel) features. The histograms are shown for SNe with available wavelength measurements interpolated to 30~days past maximum.  The blue and red dashed lines on the histograms mark the median minimum for SNe~IIb/Ib and SNe~Ic/Ic-BL at day~30, respectively. The symbols and colors are coded as in Figure~\ref{fig:pEW}.}
    \label{fig:min}
\end{figure*}

%result
The wavelengths of the absorption minima for each of the 1~\um\ and the 2~\um\ features are shown in Figure~\ref{fig:min}.
There does not appear to be a significant difference in the location of the 1~\um\ absorption minimum between the SN~IIb/Ib and SN~Ic/Ic-BL groups.
For the 2~\um\ feature, there is a slight difference between the two groups.
On average, the 2~\um\ absorption minima of SNe~Ic/Ic-BL appear at slightly redder wavelengths than those of SNe~IIb/Ib.
We convert these wavelengths to velocities for several possible ions and discuss the implications in Section~\ref{sec:discussion}.

%%%%%%%%%%%%%%%%%%%%%%%
\section{Principal Component Analysis}\label{sec:PCA}

%intro
PCA is a statistical technique that reduces the dimensionality of a data set and thus allows for many applications.
The method has been shown to be especially useful for spectroscopic data sets where there are relatively few instances of missing information \citep[e.g.,][]{Hsiao_2007, Davis_2019, Williamson_2019}. 
Here, we employed PCA to search for trends in the spectral variations that may otherwise be difficult to detect. 

%procedures
The input data are organized into a fixed rest-wavelength grid, and each spectrum is smoothed and then normalized to have the same $J$-band flux. 
All spectra were included.
The resulting principal components (PCs) are therefore heavily weighted toward a few SNe with many time-series spectra.
Standard procedures are followed for the PCA with the mean flux removed, covariance matrix calculated and diagonalized.
The PCs are then ordered according to their eigenvalues, or how much variation in the data they account for.

\begin{figure}
    \epsscale{1.2}
    \centering
    \plotone{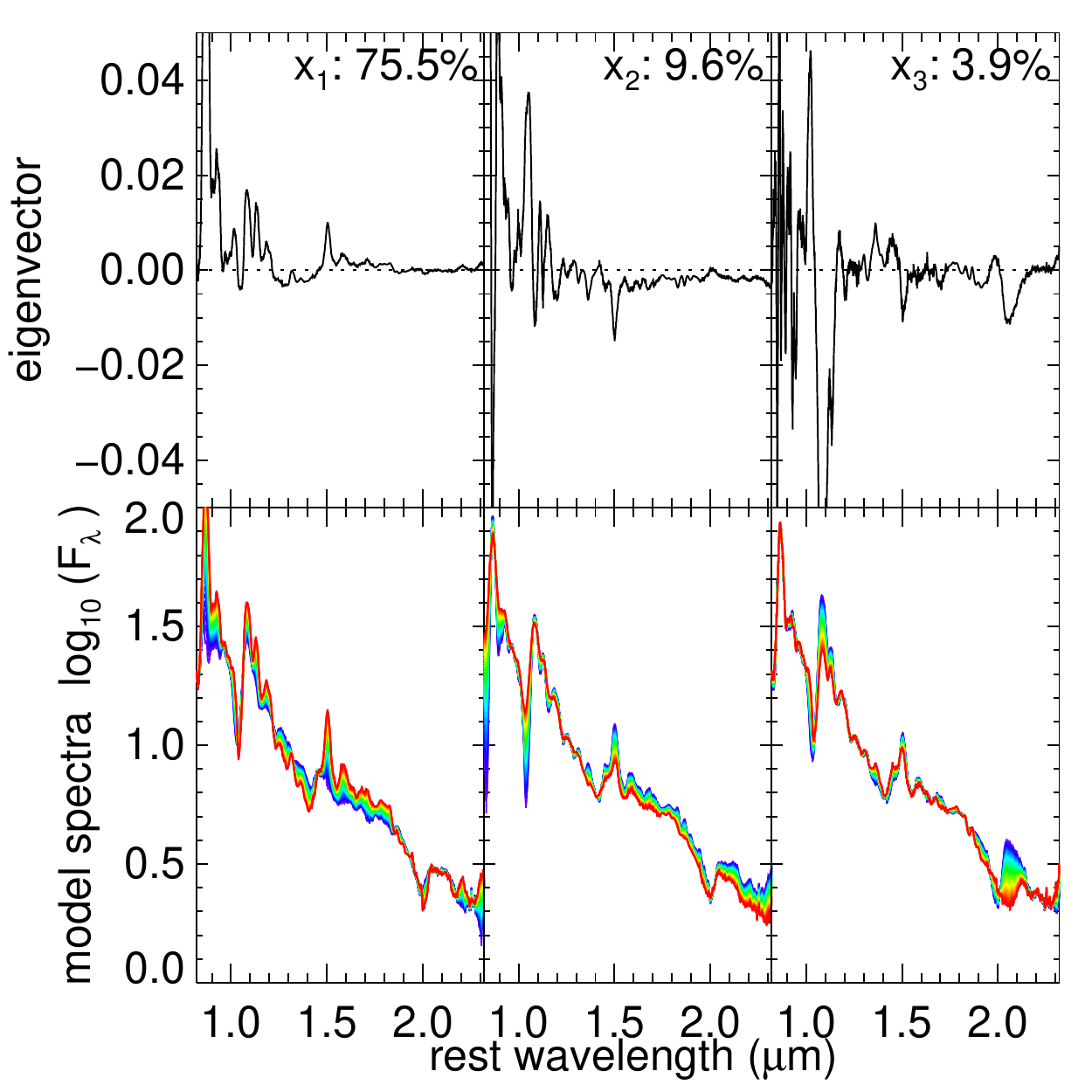}
    \caption{The first, second, and third PCs. The top panels show the eigenspectra, and the bottom panels present the PC-reconstructed spectra. The colored PC-reconstructed spectra represent the 1-$\sigma$ data variation each PC describes. The first three PCs account for 75.5\%, 9.6\%, and 3.9\% of the spectral variations, respectively. The first and second PCs describe the general time evolution in the spectral features, whereas the third PC captures the differences between the SN~IIb/Ib and SN~Ic/Ic-BL groups, mostly in the 2~\um\ region.}
    \label{fig:PCA_model}
\end{figure}

%PCs
The first four PCs account for over 90\% of the spectral variation. 
The strength of the first component alone relative to the others tells us how similar the spectra are, confirming that SESNe have broadly similar spectral features in the NIR.
The first and second PCs, accounting for $\sim$76\% and $\sim$10\% of the variation, respectively, describe the general time evolution in the spectral features (Figure~\ref{fig:PCA_model}). 
As the NIR spectra evolve, the strengths of the features increase and the velocities decrease in general (Section~\ref{sec:measurements}). 

\begin{figure}
    \epsscale{1.2}
    \centering
    \plotone{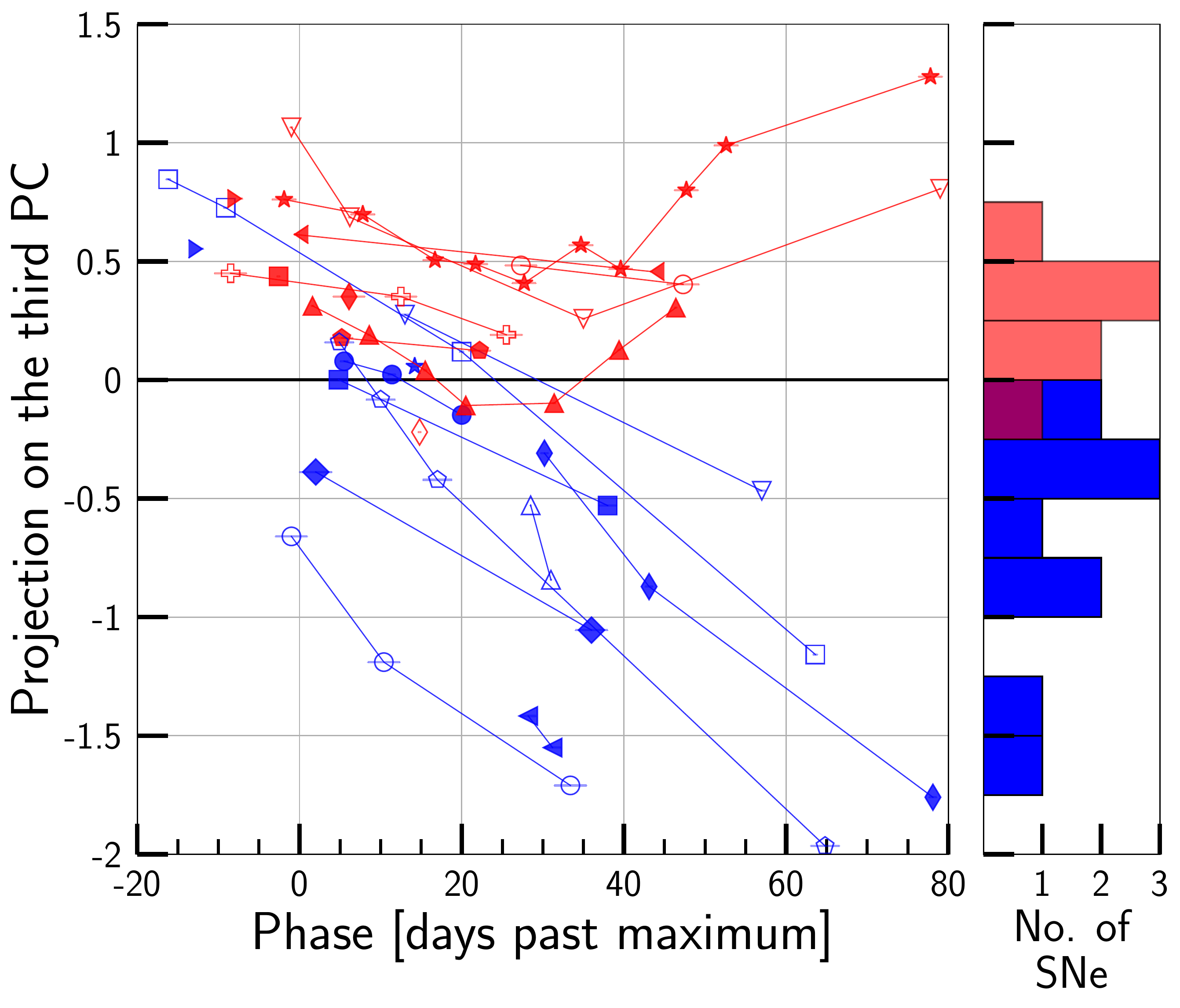}
    \caption{Projection of the spectra on the third PC shown over time. The measurements of the same SN are connected by straight lines. The He-rich Type IIb/Ib and the He-poor Ic/Ic-BL are presented in blue and red, respectively. The third PC is able to separate the He-rich and He-poor groups especially past roughly one month past maximum. The same symbols from Figure~\ref{fig:pEW} are also used here. The histograms are shown for SNe with available measurements interpolated to 30~days past maximum.}
    \label{fig:Ibc_PCA_proj}
\end{figure}

%He-rich and He-poor
The third PC only accounts for only $\sim$4\% of the overall spectral variation, but describes the differences between the NIR He-rich and He-poor groups (Section~\ref{subsec:pEW}). 
In Figure~\ref{fig:Ibc_PCA_proj}, the projection of each spectrum onto the third PC versus phase is shown. 
This PC is largely able to divide the two groups at early phases with only a few exceptions.
The groups then begin to completely diverge starting at roughly one month past maximum.
The divergence of the NIR spectral properties again demonstrates that the NIR He-rich and He-poor groups correspond almost perfectly with the optical SN~IIb/Ib and SN~Ic/Ic-BL groups.
It bears repeating here that the classifications are entirely based on optical spectra near maximum light. 

\begin{figure}
    \epsscale{1.15}
    \centering
    \plotone{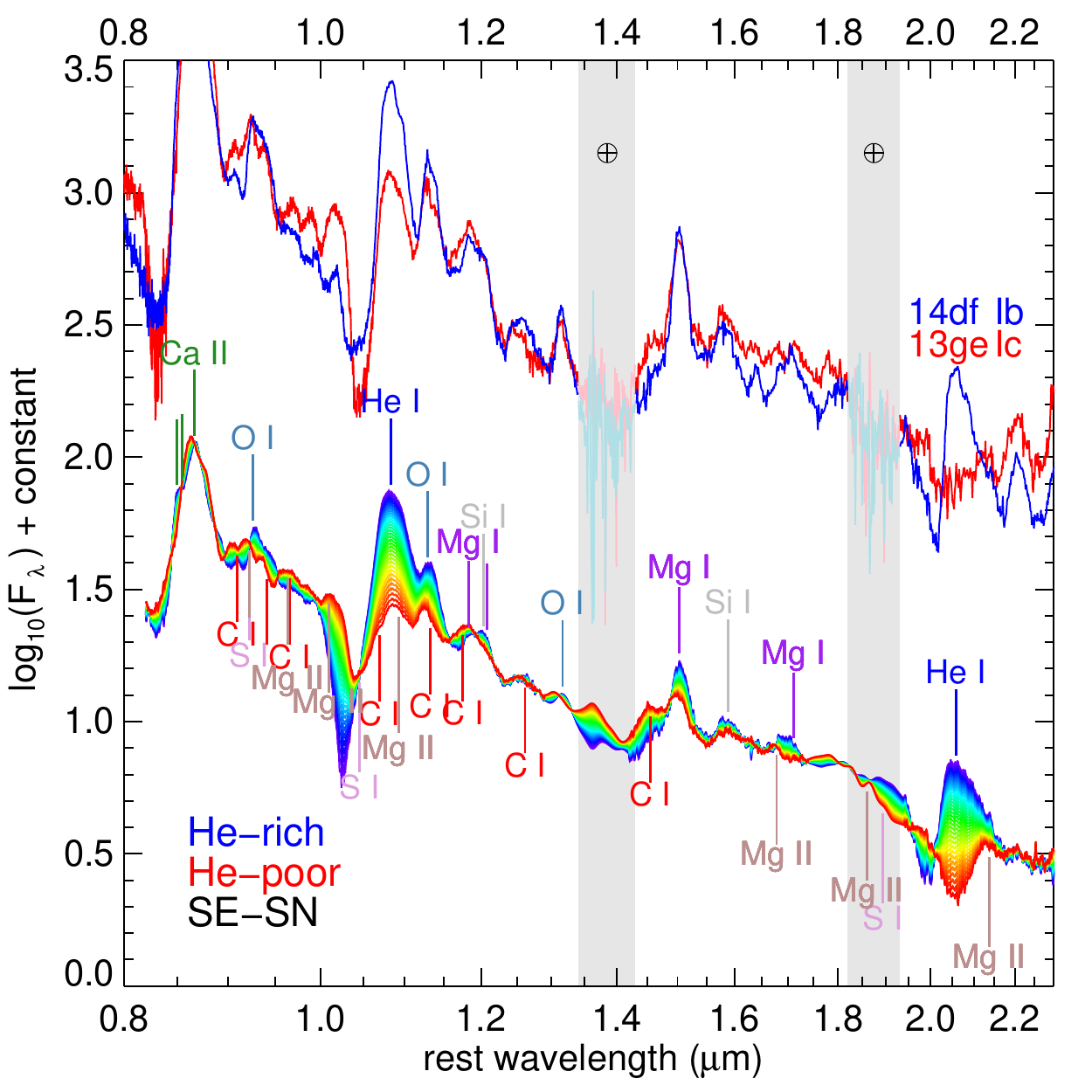}
    \caption{Line identifications of the NIR He-rich and He-poor groups using the third PC. The PC-reconstructed spectra are presented on the bottom representing the spread in the projections shown in Figure~\ref{fig:Ibc_PCA_proj} (the color range from blue to red represents projections from $-2.0$ to $1.3$). The vertical lines mark the laboratory rest wavelengths of the labeled ions. The 2~\um\ region shows the largest difference in profile shapes between the two groups. The spectra of SNe~2014df (Ib) and 2013ge (Ic) at late phases are also shown as examples of the He-rich and He-poor groups, respectively.}
    \label{fig:Ibc_PCA_lineID}
\end{figure}

%line ID in 2um region
Taking a closer look at the differences in the NIR spectroscopic properties between the He-rich and He-poor groups, Figure~\ref{fig:Ibc_PCA_lineID} plots the spectral variations reconstructed by the third PC.
Note that the PCs have been smoothed in the regions of heavy telluric absorptions, and those regions are marked in Figure~\ref{fig:Ibc_PCA_lineID}.
The most discernible differences are observed near the 2~\um\ region where the \ion{He}{1} \lam2.0581~\um\ line resides, confirming the results from Section~\ref{subsec:pEW}. 
For the He-poor group, while the \ion{He}{1} \lam2.0581~\um\ line may be weak, the \ion{Mg}{2} \lam2.1369~\um\ may be present further to the red (also seen in Figure~\ref{fig:2micron_pl}).
This suggests that the 1~\um\ feature may be partially attributable to \ion{Mg}{2} \lam1.0927~\um\ for the He-poor group.

%Other line ID
Since the strong feature near 1~\um\ may be attributed to several strong lines such as \ion{C}{1}, \ion{He}{1}, and \ion{Mg}{2} (Section~\ref{sec:lineid}), the 2~\um\ region may indeed be the best region to determine whether \ion{He}{1} is present in a NIR spectrum.
More subtly, the third PC shows more prevalent \ion{He}{1} and \ion{Mg}{1} features in He-rich SNe and more prevalent \ion{C}{1} and \ion{Mg}{2} features in He-poor SNe. 
These ions are identified using the emission peaks of P Cygni profiles of multiple lines in the PC-reconstructed spectra in Figure~\ref{fig:Ibc_PCA_lineID}.
In individual spectra, the identifications may be less obvious or individual features may disappear with spectral evolution. 

%spectral template
Using the resulting PCs and their correlations, we construct the NIR spectral templates for SESNe that may be useful for a variety of applications. 
First, the time evolution of the projections on each of the four PCs is specified for each SN type.
This is an imperfect process, as there exist variations within each spectral type (see Figure~\ref{fig:Ibc_PCA_proj}). 
SNe~IIb and Ib have indistinguishable projections for the first four PCs, so one spectral template is built for both. 
This further supports our assessment that the H Paschen and Brackett series in the NIR are weak for SNe~IIb compared to SNe~II.

%%%%%%%%%%%%%%%%%%%%
\section{Discussion}\label{sec:discussion}

The main focus of this work is to identify the ions responsible for the prominent features near 1~\um\ and 2~\um. 
Specifically, whether they correspond to the \ion{He}{1} \lam1.0830~\um\ and \lam2.0581~\um\ lines is of great interest especially for SNe~Ic.
There are several strong lines that may contribute to the 1~\um\ feature (Section~\ref{sec:lineid}), complicating the analysis.
Thus, we use several methods in this section, in an attempt to piece together a coherent picture and identify the most likely dominant species for each SN group. 

\subsection{Line identifications through observables}
\label{sec:lineid_obs}

%1um feature
\paragraph{The 1~\um\ feature} 
The strongest \ion{He}{1} line at both the optical and the NIR wavelengths is the \ion{He}{1} \lam1.0830~\um\ line.
Nearly all of the spectra in our sample have a strong absorption near 1~\um\ making it the natural place to begin our investigation.
The profile shape of this feature varies widely, and a two-Gaussian function provides adequate fits in most instances (Section~\ref{subsec:velocity}).
This does not automatically assume that there are two lines contributing to the feature, as lines formed in large optical depth can produce a similar shape.
Nonetheless, we explore the possibility that the profile is formed by multiple ions.

\begin{figure*}
    \epsscale{1.1}
    \centering
    \plotone{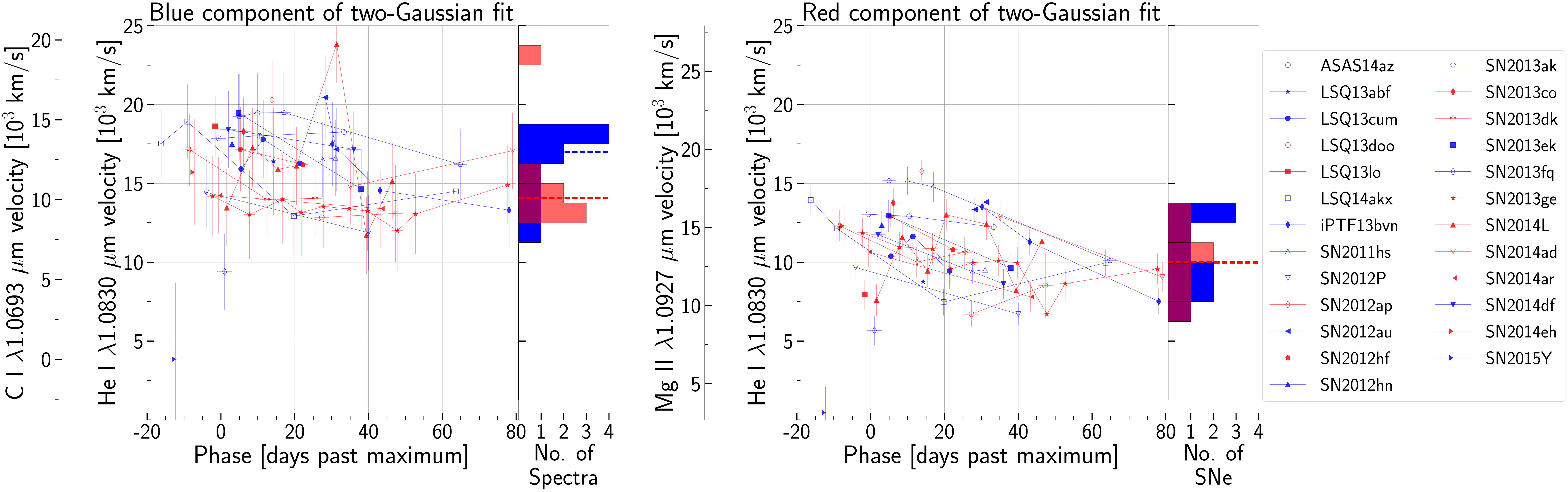}
    \caption{Velocity of the absorption feature at 1~\um, assuming that the two-Gaussian fitting function represents two separate ions. The SN~IIb/Ib and SN~Ic/Ic-BL groups are shown in blue and red, respectively, with filled symbols representing Type Ib/Ic and open symbols representing Type IIb/Ic-BL. The left panel shows the blue component of the two-Gaussian function. The right panel shows the red component. The assumed ion is indicated for each panel on the y-axis. The histograms are shown for SNe with available velocity measurements interpolated to 30~days past maximum. The blue and red dashed lines on the histograms mark the median velocity for SNe~IIb/Ib and SNe~Ic/Ic-BL at day~30, respectively.}
    \label{fig:velocities_comps}
\end{figure*}

%velocities of two components
In Figure~\ref{fig:velocities_comps}, the blue and red components of the two-Gaussian functions from the 1~\um\ profile fits are examined separately, and the velocities are shown assuming several possible ions.
Three strong lines in this region were considered: \ion{C}{1} \lam1.0693~\um, \ion{He}{1} \lam1.0830~\um, and \ion{Mg}{2} \lam1.0927~\um.
The velocity of the blue component is shown assuming that the main contributor is \ion{C}{1} or \ion{He}{1}; the velocity of the red component is shown assuming that the main contributor is \ion{He}{1} or \ion{Mg}{2}.
All possibilities explored result in reasonable ranges of velocities, except for a few outliers. 
Furthermore, on average, we observe that the velocities are decreasing over time. 
This agrees with the study done by \citet{Liu_2016} on the optical spectra of SESNe. 
It is also worth noting that \citet{Liu_2016} showed the biggest difference between SNe~IIb/Ib and SNe~Ic can be found in the strength of the optical \ion{O}{1} line. 
They also found that SNe~Ic show higher velocities for both \ion{O}{1} and \ion{Fe}{2} optical lines.

%velocity ranges
As shown in Figure \ref{fig:velocities_comps}, assuming \ion{He}{1} for the blue component, velocities of the 1~\um\ feature range from $12{,}500$ to $20{,}000$~\kms.
For the blue component, He-poor SNe show slightly slower velocities on average than He-rich SNe.
As SNe~Ic are known to be more energetic explosions \citep{Mazzali_2017}, it is not likely that the 1~\um\ feature is formed by the same ions for both He-poor and He-rich SNe.
On the other hand, assuming \ion{He}{1} for the red component results in a velocity range of $7{,}000$ to $15{,}000$~\kms, and a distinction is not seen between the two groups. 
Assuming \ion{C}{1} for the blue component and assuming \ion{Mg}{2} for the red component, the resulting velocity ranges are $8{,}000-16{,}000$~\kms\ and $10{,}000-17{,}500$~\kms, respectively.

%combining two components
The combination of a blue \ion{He}{1} and a red \ion{Mg}{2} component would mean the Mg layer is situated only slightly below the He layer with a significant overlap. 
Noting that this is true in a spherically symmetric explosion with concentric layers \citep[e.g.][]{Bersten_2012}.
The combination of a blue \ion{C}{1} and a red \ion{He}{1} component would mean the C layer is situated slightly above the He layer again with a significant overlap. 
This scenario would require contrived mixing and not likely to be true.
The combination of a photospheric and high-velocity He component or a single He line formed in large optical depth are also possibilities. 
There are no distinct differences between the He-rich and He-poor groups in the 1~\um\ region.

\begin{figure}
    \epsscale{1.1}
    \centering
    \plotone{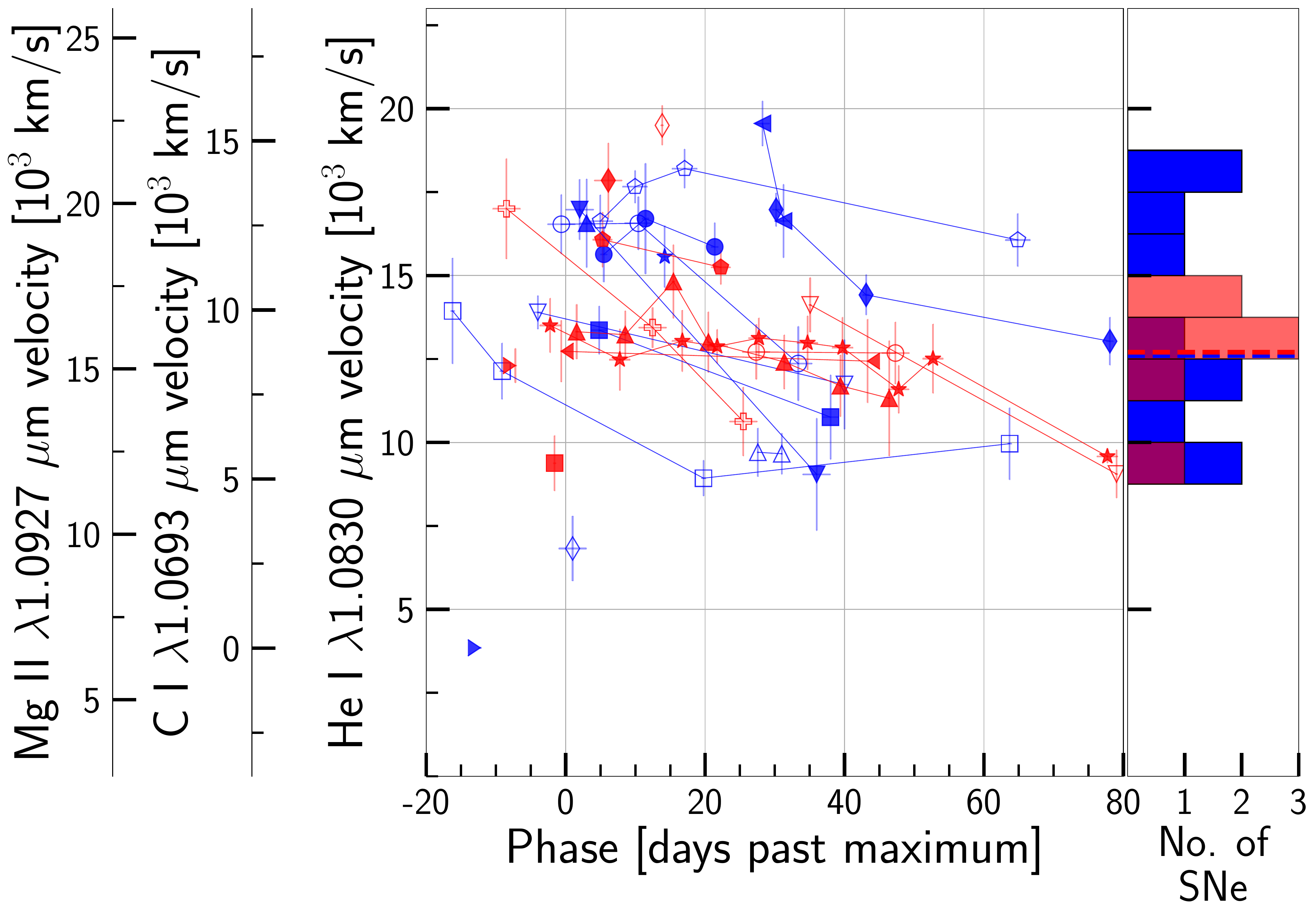}
    \caption{Velocity of the absorption feature at 1~\um, assuming that the two-Gaussian fitting function represents a single ion. The assumed ion is indicated for each panel on the y-axis. The same symbols from Figure~\ref{fig:velocities_comps} are used here. The histograms are shown for SNe with available velocity measurements interpolated to 30~days past maximum.  The blue and red dashed lines on the histograms mark the median velocity for SNe~IIb/Ib and SNe~Ic/Ic-BL at day~30, respectively.}
    \label{fig:velocities}
\end{figure}

%velocities of a single profile
In the case that the entire 1~\um\ profile is formed by a single line, we then plot the velocities measured from the profile minima of the entire two-Gaussian fit. 
Again, three possibilities are considered: \ion{C}{1} \lam1.0693~\um, \ion{He}{1} \lam1.0830~\um, and \ion{Mg}{2} \lam1.0927~\um\ (Figure~\ref{fig:velocities}).
The resulting velocities also occupy reasonable ranges of $10{,}000-20{,}000$~\kms\ for \ion{He}{1}, $5{,}000-15{,}000$~\kms\ for \ion{C}{1}, and $12{,}000-22{,}000$~\kms\ for \ion{Mg}{2}.
The velocities would be slightly high for \ion{Mg}{2}, with some SNe reaching above $20{,}000$~\kms, making it unlikely that this feature is formed solely by \ion{Mg}{2}.
There are also no distinct differences between the He-rich and He-poor groups in this exercise.

\begin{figure}
    \epsscale{1.1}
    \centering
    \plotone{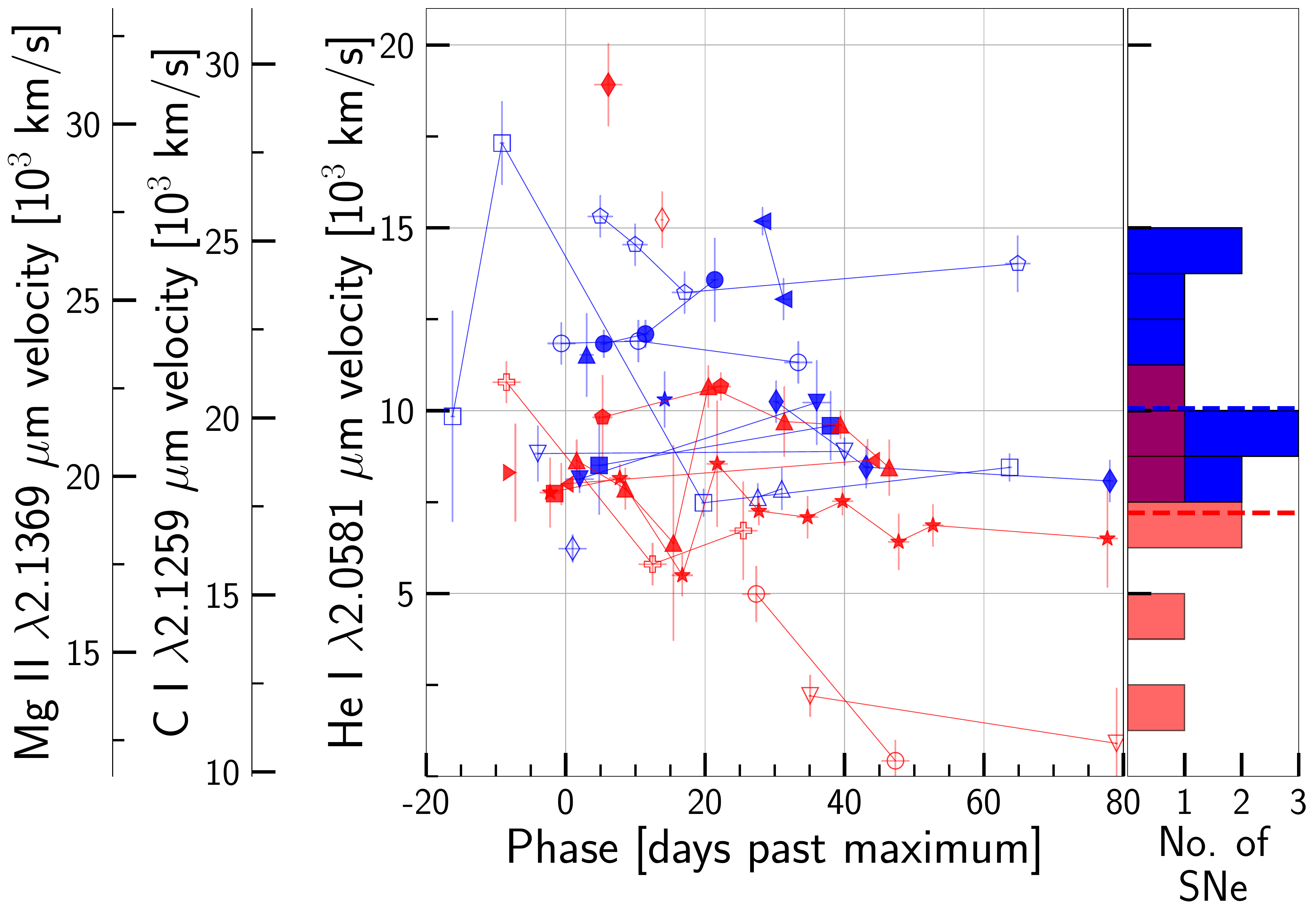}
    \caption{Velocity of the absorption feature at 2~\um\ measured by fitting a two-Gaussian function. The assumed ion is indicated for each panel on the y-axis. The same symbols from Figure~\ref{fig:velocities_comps} are used here. The histograms are shown for SNe with available velocity measurements interpolated to 30~days past maximum.  The blue and red dashed lines on the histograms mark the median velocity for SNe~IIb/Ib and SNe~Ic/Ic-BL at day~30, respectively.}
    \label{fig:velocities_2}
\end{figure}

%2um feature
\paragraph{The 2~\um\ feature} Although the overlaps between strong lines are less severe in the 2~\um\ region, the profile shapes of the 2~\um\ feature are not always well-described by a single Gaussian function.
A two-Gaussian fit was then chosen to determine the radial velocity shift. 
Note that for most spectra, there are two distinct features detected in the range of $1.95-2.15$~\um. 
For each of these cases, the feature on the blue side is our target for fitting with the two-Gaussian function (dubbed the 2~\um\ feature), since it has been suggested in Section~\ref{sec:PCA} that the feature on the red side may be attributed to \ion{Mg}{2} \lam2.1369~\um.
In the case where the features are blended and indistinct, the velocity was then measured from the minimum of the entire profile. 
The results are represented in Figure~\ref{fig:velocities_2}.

%line ID
Three lines are considered here for the identity of the 2~\um\ feature: \ion{He}{1} \lam2.0581~\um, \ion{C}{1} \lam2.1259~\um, and \ion{Mg}{2} \lam2.1369~\um\ (Figure~\ref{fig:velocities_2}).
Assuming that the 2~\um\ feature is mainly formed by \ion{He}{1}, the velocity ranges from approximately $5{,}000$ to $15{,}000$~\kms.
Assuming that the same feature is formed by \ion{C}{1} and \ion{Mg}{2} results in velocity ranges of $15{,}000-25{,}000$~\kms\ and $17{,}000-27{,}000$~\kms, respectively.
Such high velocities for \ion{C}{1} and \ion{Mg}{2} are unlikely, especially when a feature to the red, often distinct from the 2~\um\ feature, yields reasonable velocity ranges of $3{,}000-8{,}000$ and $5{,}000-10{,}000$~\kms\ for \ion{C}{1} and \ion{Mg}{2}, respectively.

%overall properties
Note that the \ion{He}{1} velocities derived from the 2~\um\ feature are systematically lower than that indicated by the 1~\um\ feature.
This apparent difference is expected if the 1~\um\ feature is formed at large optical depths, that tend to push the minimum toward the blue (Section~\ref{subsec:lineid_mod}). 
Similar to the result from the 1~\um\ feature, the 2~\um\ feature tends to be at redder wavelengths indicating lower velocities on average in the He-poor group compared to the He-rich group.
Considering that SNe~Ic/Ic-BL are on average more energetic events than SNe~IIb/Ib, the 2~\um\ feature may not have been formed by the same lines in both groups.

%2um feature: PCA and pEW
Both the pEW measurements (Section~\ref{subsec:pEW}) and the PCA (Section~\ref{sec:PCA}) point to the 2~\um\ region as the ideal place to detect the presence of He.
These analyses were able to distinguish NIR He-rich and He-poor groups that almost perfectly correspond to the optical IIb/Ib and Ic/Ic-BL designations, respectively.
The examinations of the 2~\um\ observations also show that He-rich and He-poor SNe have distinct properties.
Thus, based on our observations, the division between the two groups is not an arbitrary one along a continuous sequence.
The NIR observations then lend credence to the optical Ib/Ic division.

\subsection{Search for residual He in He-poor SNe}\label{sec:residual_He}

\begin{figure}
    \epsscale{1.15}
    \centering
    \plotone{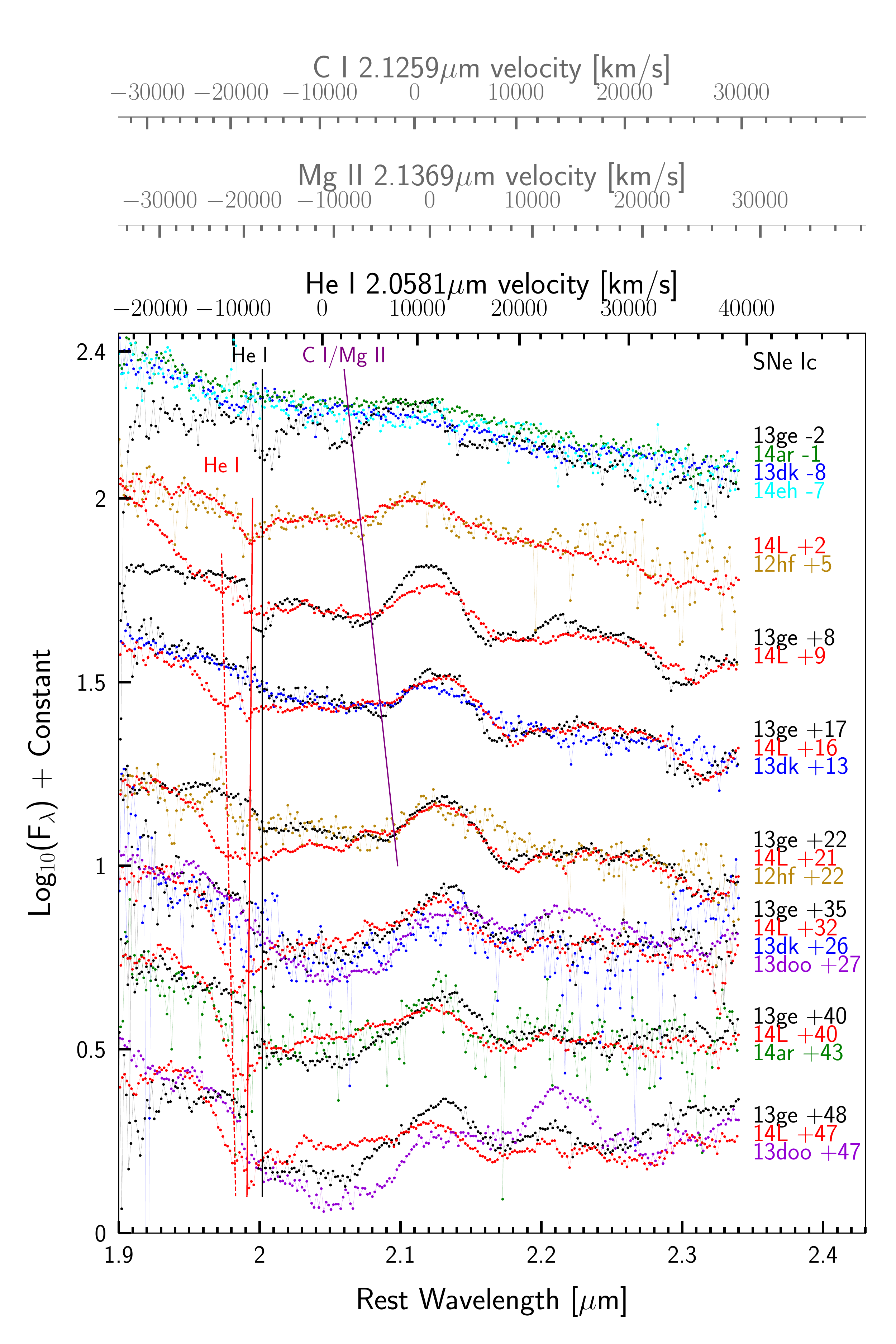}
    \caption{Spectral evolution of SNe~Ic in the 2~\um\ region. The top x-axis shows the velocity scale of the \ion{He}{1} \lam2.0581~\um, \ion{Mg}{2} \lam2.1369~\um, and \ion{C}{1} \lam2.1259~\um\ lines. The vertical lines indicates possible \ion{He}{1} evolution for SN~2013ge (black) and SN~2014L (red). There may be two components for SN~2014L (see text for details). The purple lines indicates a possible \ion{C}{1}/\ion{Mg}{2} feature. SNe Ic-BL have been excluded from this plot. The spectra of SN~2013co and LSQ13lo have also been excluded due to their low S/N ratios in the region.}
    \label{fig:2micron_pl}
\end{figure}

%2um feature of SNe~Ic
We now turn to the question of whether He-poor SNe do in fact harbor trace amounts of He.
In Figure~\ref{fig:2micron_pl}, the 2~\um\ region is presented for all the SN~Ic spectra in our sample with high enough S/N ratio in the $K$ band.
A significant fraction of the He-poor spectra in the sample show weak absorption features near 2~\um.
In particular, SN~2013ge and SN~2014L, two of the best-sampled He-poor SNe, show varied time evolution and the persistence of this feature.
\citet{Drout_2016} identified the weak feature in SN~2013ge as \ion{He}{1} \lam2.0581~\um\ and proposed that the feature persists out to $+40$~d past maximum (black vertical line in Figure~\ref{fig:2micron_pl}).
It was then classified as a SN~Ib/c by \citet{Drout_2016} using the combination of the optical and NIR He lines.
The evolution of \ion{He}{1} \lam2.0581~\um\ in SN~2014L is more complex, possibly showing two components emerging at different epochs (dashed and solid red vertical lines in Figure~\ref{fig:2micron_pl}).
The component indicated by the red dashed line in Figure~\ref{fig:2micron_pl} emerges later and continues to strengthen at late epochs.
Indeed, this component may be formed by other ions, and the \ion{He}{1} features in both SN~2013ge and SN~2014L may only persist to $1-2$ weeks past maximum rather than $>1$ month as previously proposed. 
This could be due to the He-layers becoming optically thin by these epochs. 

%statistics
Among the SNe~Ic in our sample, SN~2014L, SN~2013ge, and SN~2012hf show evidence of trace amounts of He through the weak \ion{He}{1} \lam2.0581~\um\ absorption. 
However, the early and high S/N ratio spectra of SN~2014ar, SN~2013dk, and SN~2014eh can rule out the existence of \ion{He}{1} \lam2.0581~\um\ absorption.
LSQ13doo does not have an early enough spectrum for this line to be ruled out.
To summarize, even though the \ion{He}{1} \lam2.0581~\um\ absorption in SNe~Ic is much weaker compared to SNe~IIb/Ib, approximately half of the SNe~Ic in our sample show a weak absorption.
Although this is a significant fraction, we caution that our sample is in no way statistically complete or representative of the population in nature.
As mentioned previously, the distinct feature to the red side near 2.1~\um\ and well separated from the \ion{He}{1} complex, may be attributed to \ion{C}{1} \lam2.1259~\um\ and/or \ion{Mg}{2} \lam2.1369~\um\ (purple vertical line in Figure~\ref{fig:2micron_pl}). 

\subsection{Line identifications through models} 
\label{subsec:lineid_mod}

\begin{figure}
    \epsscale{1.2}
    \centering
    \plotone{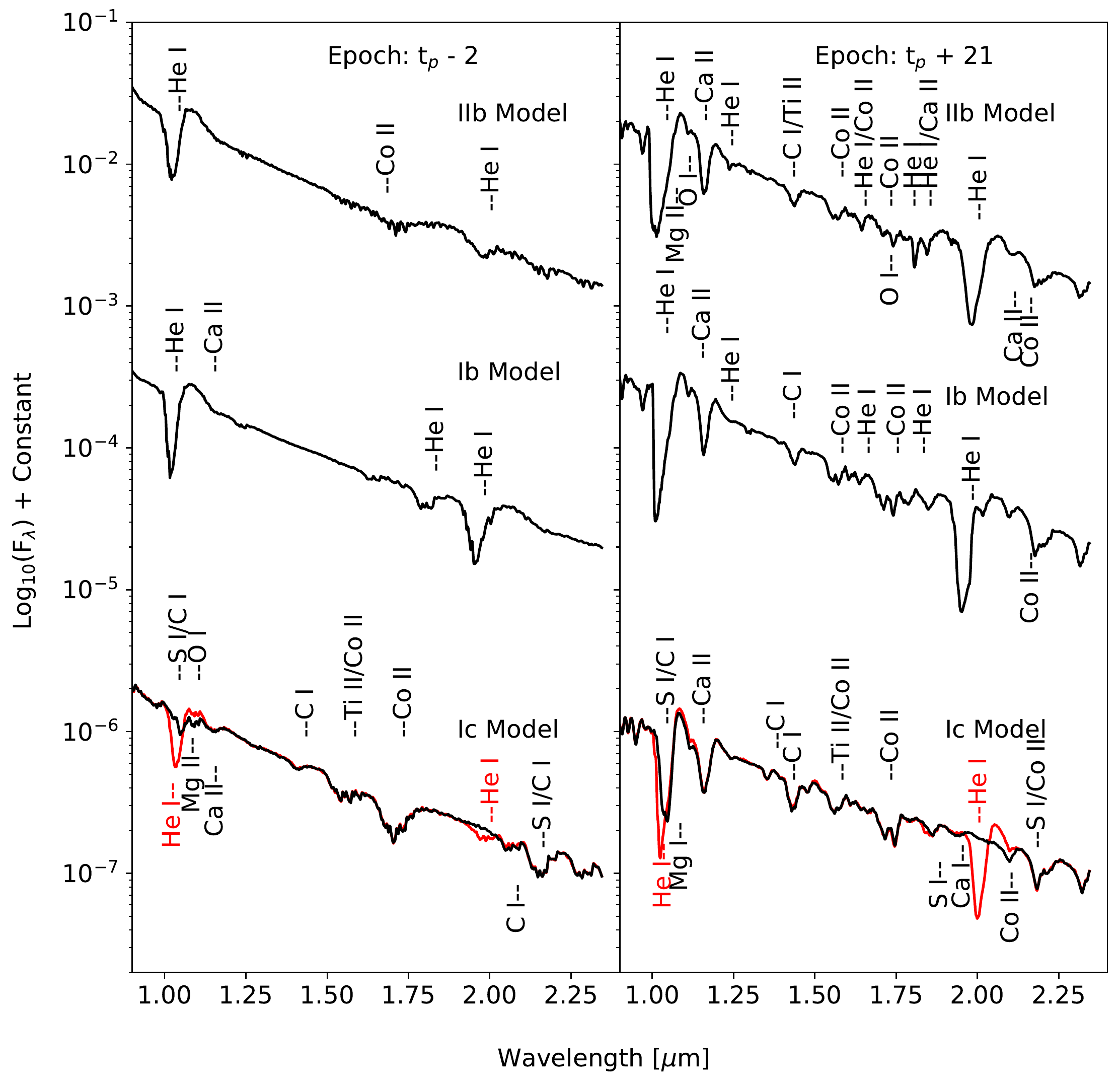}
     \caption{NIR line identifications based on a subset of synthetic spectra from \cite{Teffs_2020_b}. They are presented at two epochs: $2$~days prior and $+21$~days past peak luminosity, plotted in the left and right panels, respectively. The SN~IIb, Ib, and Ic model spectra are shown at each epoch. For the SN~Ic model, the red spectra model has 0.02~M$_{\sun}$ of He mixed evenly through the C/O core while the black has no He mass in the model. The ions responsible for the absorption features are labeled. See text for the model properties.
     }
    \label{fig:lid}
\end{figure}

In this subsection, we use the models of \citet{Teffs_2020_b} to compare with observations and guide us in line identifications. 
For alternative models see \citet{Dessart_2015, Dessart_2020}.
Figure~\ref{fig:lid} presents a selection of line identifications for a set of synthetic SNe~IIb/Ib/Ic spectra at two epochs: 2 days prior and 21 days past peak luminosity.
Details regarding the explosion models are given in \citet{Teffs_2020_b}.

%IIb
For all models, we start with a 22~M$_{\sun}$ progenitor stripped of H/He of varying degrees prior to collapse. 
The SN~IIb model is exploded at a final energy of 3 foe and ejects 5.8~M$_{\sun}$ of material comprised of an atmosphere of 0.1~M$_{\sun}$ of H and 1.3~M$_{\sun}$ of He above an evenly mixed C/O rich core.
These mixing approximations result in the early spectral features forming primarily in the H/He-rich regions, thus containing minimal contributions from other elements such as O and Ca.
By 21 days past maximum, the photosphere has receded into the C/O rich core, producing several of the C, O, Mg, and Fe-group lines as identified in Section~\ref{sec:lineid}. 
The rich He shell is still the dominant line forming region, producing easily identifiable 1 and 2 \um\ lines.

%Ib
The SN~Ib model is also exploded at a final energy of 3 foe, and ejects 5.3~M$_{\sun}$ of material with 1.0~M$_{\sun}$ of He.
This model uses a similar mixing approximation as the IIb model, with a He rich shell above a mixed C/O core and was shown in \citet{Teffs_2020_b} to match several SNe~Ib in the optical. 
Similar to the IIb model, at 2 days prior to peak luminosity, the photosphere is still in the He rich outer shell. 
The apparent strength of the \ion{He}{1} $\lambda$2.0581\um\ line is stronger in the Ib model compared to that in the IIb, due to the relative abundances above each photosphere of H/He.
By 21 days past maximum, the SN~IIb and Ib synthetic spectra show few differences, as shown in the observed spectrum in the previous sections. 
The lack of the Paschen series in the NIR for our H-poor models results in effectively both the IIb/Ib having a C/O rich core below a He shell.

%Ic
The SN~Ic model is also exploded at a final energy of 3 foe, and ejects 4.2~M$_{\sun}$ of material with no He and is thoroughly mixed. 
A second SN~Ic model, that shows minimal to no optical features, is included with 0.02~M$_{\sun}$ of He mixed throughout the ejecta.
This model has a $\sim$ 20\% reduction in luminosity across all epochs when compared to the model in \cite{Teffs_2020_b} to fit the NIR features in SN~2014L.
The spectrum of the He-free SN~Ic model at 2 days before maximum shows no dominant line near 1\um. 
Instead, the region is comprised of three weaker \ion{S}{1}, \ion{C}{1}, and \ion{Mg}{2} lines that become significantly stronger with time. 
In the model with trace amounts of He, we find a strong absorption feature due to \ion{He}{1}.
Note however that the vast majority of SNe~Ic in our observed sample exhibit the strong 1~\um\ absorption feature from the earliest phases.
Based on the mixing approximation used, this may not place enough C, S, or Mg at higher velocities that would generate a dominant, non-\ion{He}{1} feature at 1\um.

Given the late-time dominance of the \ion{S}{1} and \ion{C}{1} lines of the 1~\um\ feature in the models, modification of outer abundance structure may still reproduce the observed early feature without the need for He.
Additionally, the 1\um\ feature in the later epoch is dominant in both models with little difference between the models with and without He.
As the photosphere recedes in the C/O core for both epochs, the spectra show much stronger \ion{Ca}{2} and \ion{C}{1} lines than in the SN~Ib model.
The strong \ion{Ca}{2} \lam1.1839~\um\ and \lam1.1950~\um\ lines are also present in the observed SN~Ic spectra.
Consistent with observations, the largest difference between the He-rich SN~IIb/Ib and the He-poor SN~Ic models is in the 2~\um\ feature where the He-rich models are dominated by the \ion{He}{1} \lam2.0581~\um\ line.

%Co II
One prevalent ion in all of the synthetic spectra is \ion{Co}{2}.
The features formed by \ion{Co}{2} may be a consequence of the mixing approximations for $^{56}$Ni chosen in these models.
It is unclear whether they are present in the observed spectra.
The $^{56}$Ni distribution in the SN~IIb/Ib models assumes a central mass of $^{56}$Ni that has a running boxcar average across the ejecta, resulting in a relatively smooth decrease in the abundance of $^{56}$Ni outside the core. 
The SN~Ic model, however, has an even distribution of $^{56}$Ni throughout the ejecta and the over-abundance of $^{56}$Ni in the outermost region may result in stronger \ion{Co}{2} features in the synthetic spectra.

\begin{figure*}
    \epsscale{1.1}
    \centering
    \plottwo{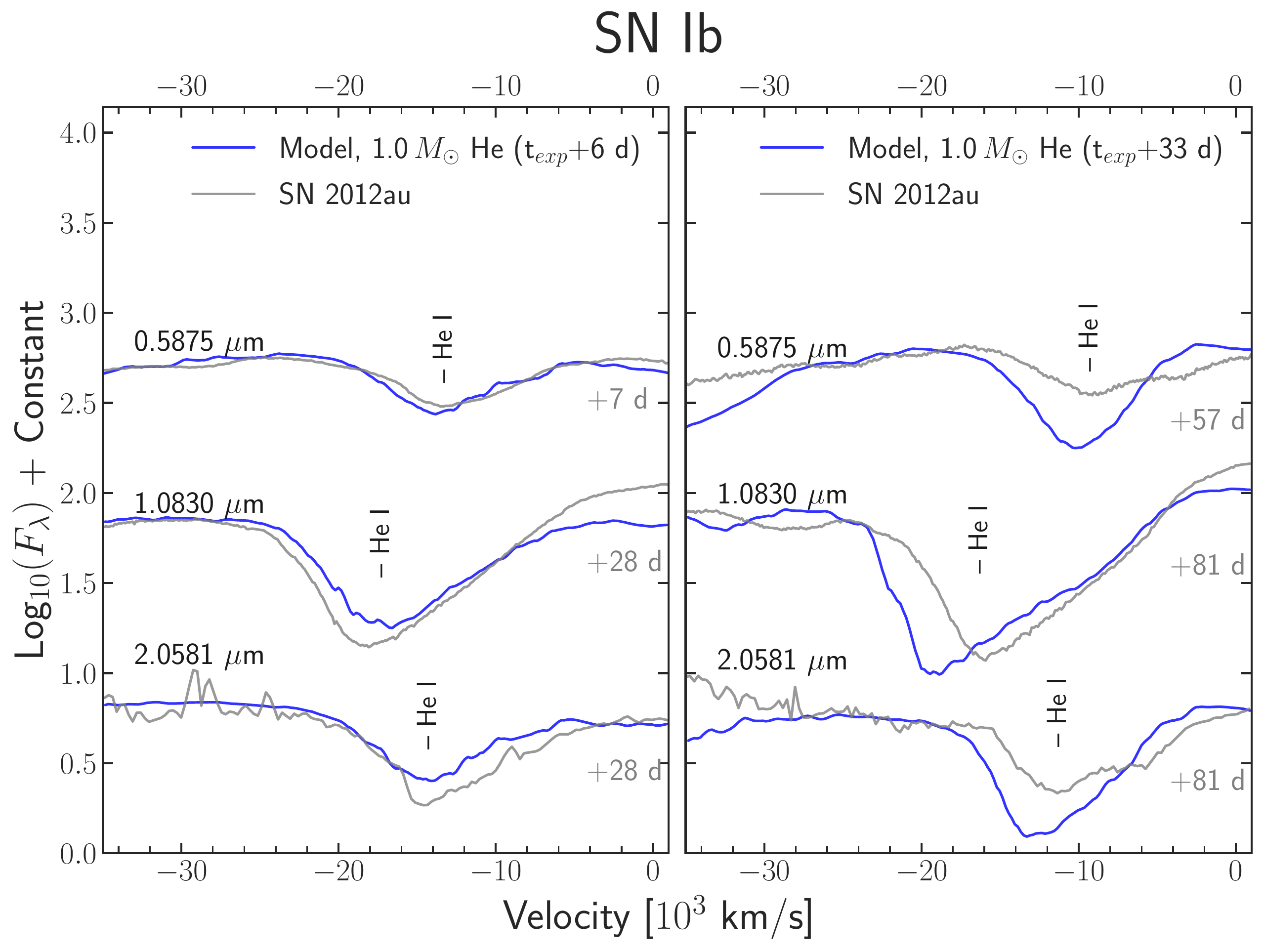}{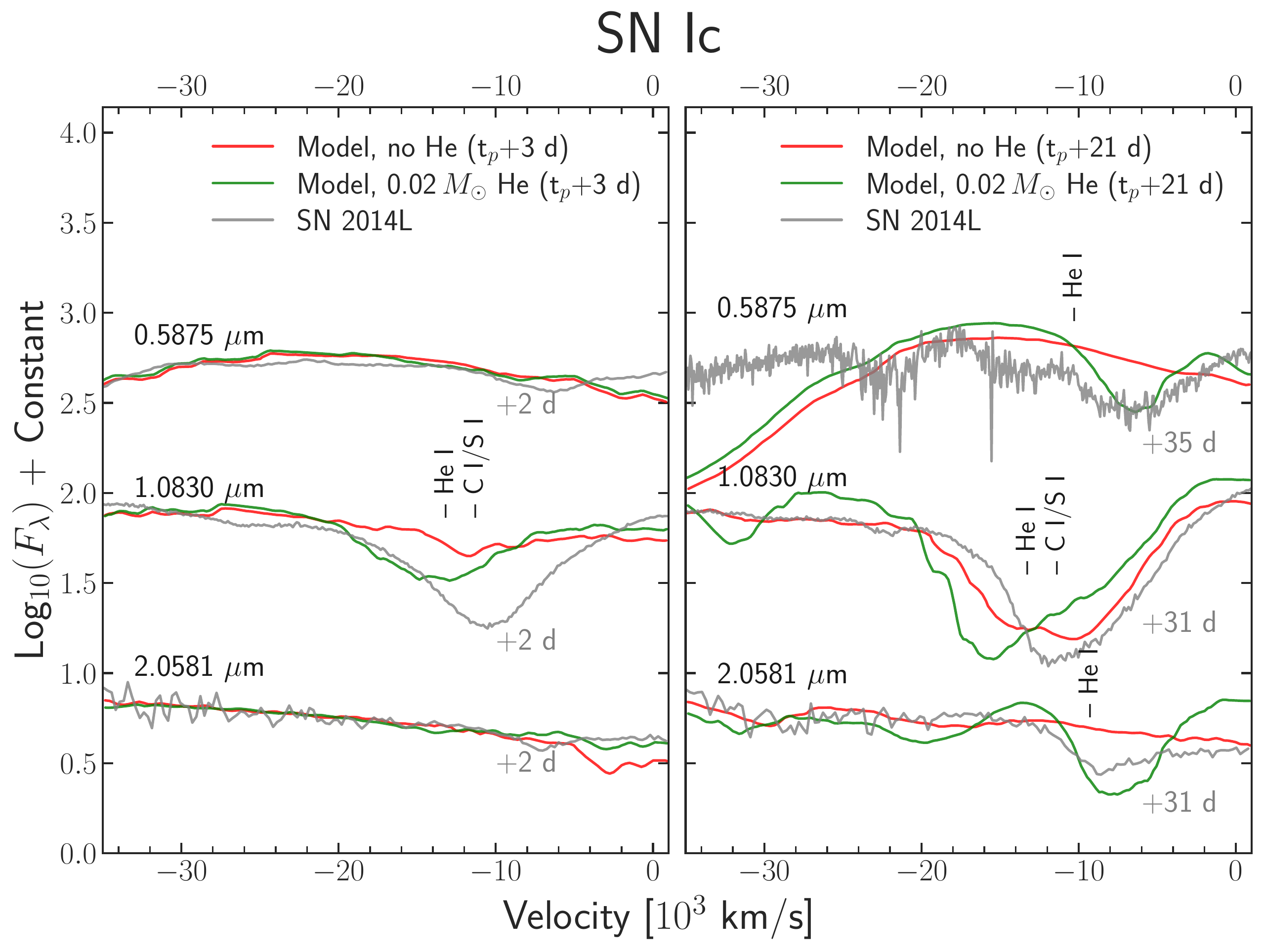}
    \caption{Comparison between observed spectra and the synthetic model spectra of \citet{Teffs_2020_b} in the velocity space of \ion{He}{1}.
    The observed optical and NIR spectra of Type Ib SN~2012au (left panel) and Type Ic SN~2014L (right panel) are shown in two epochs.
    They are compared to the SN~Ib model (left panel) and the SN~Ic model with a contribution of 0.02 M$_{\sun}$ of \ion{He}{1} included (right panel).}
    \label{fig:lid5}
\end{figure*}

%He-rich comparison
In the left panel of Figure~\ref{fig:lid5}, the observed spectra of Type Ib SN~2012au and the synthetic spectra of the SN~Ib model described above are compared in the velocity space of \ion{He}{1} at two epochs and for the three strongest \ion{He}{1} lines: \lam0.5875~\um, \lam1.0830~\um, and \lam2.0581~\um.
The model can largely reproduce the observed velocity and profile shape.
Minor mismatches at later phases may be attributed to the differences in epochs, densities, and ionization/excitation regimes.
Most notably, both the 1~\um\ and 2~\um\ features can easily be dominated by minimal to moderate amounts of He, resulting in broadened wings. 
This explains the observed asymmetric profile shapes and highlights the difficulty in determining the velocity using these features.

%He-poor comparison
In the right panel of Figure~\ref{fig:lid5}, the observed spectra of Type Ic SN~2014L are compared to the synthetic spectra of the SN~Ic model in the velocity space of \ion{He}{1}.
For this comparison, 0.02~M$_{\sun}$ of He is included in the SN~Ic model.
If He contributes to all three features in the three wavelength regions, then the velocity appears to be consistent.
Note that it is always difficult to rule out Na~I~D being the dominant species in the optical region. 
The match in the 2~\um\ region at the later epoch is substantially improved by the addition of the trace amount of He.
This result is consistent with our finding that some SNe~Ic in our sample harbor small amounts of He (Section~\ref{sec:residual_He}).
The optical feature in SN~2014L may be improved simply by introducing Na into the model, which is an element not included in the models in \citet{Teffs_2020_a}.
However, caution should be taken from these comparisons. 
The flux for both synthetic and observed SNe are scaled or normalized, and are not given in absolute flux values.
The luminosity and photospheric temperature required to reproduce the observed flux level in the SN may not be the same parameters used to calculate the synthetic spectra.
Furthermore, \citet{Teffs_2020_a} showed that for SNe~Ic, the completely mixed C/O core was successful in reproducing the observed behavior of several observed SNe. 
However, SNe~Ic directly modeled using the abundance tomography method, such as SN~2004aw \citep{Mazzali_2017}, SN~1994I \citep{Sauer_2006}, and SN~2017ein \citep{Teffs_2021}, require varied abundance distributions in order to reproduce the observed spectra.
A proper abundance stratification model following the methods in 
\cite{Stehle_2005} and \cite{Ashall_2019} would be required to determine if SN~2014L has He in the ejecta but is beyond the scope of this work.

%%%%
\section{Conclusions}
\label{sec:summary}
%dataset
We have presented the largest NIR spectroscopic data set of SESNe to date, which encompasses 75 NIR spectra of 34 \SE\ observed by CSP-II between 2011 and 2015.
The spectra past 80 days relative to maximum are excluded from this paper and will be discussed in a following paper that mostly focuses on the detection of CO. 

Our main findings for this work are summarized as follows:

% 1~\um\
1) All spectra, regardless of their optical classification, show a broad and strong absorption profile at around 1~\um.
There are multiple strong lines in this region that may be responsible for this feature, complicating the analysis.
These include: \ion{He}{1} \lam1.0830~\um, \ion{C}{1} \lam1.0693~\um\, and \ion{Mg}{2} \lam1.0927~\um.
In SNe~Ib and IIb, the \ion{He}{1} line is likely to be the main contributor to this feature. 
In SNe~Ic and Ic-BL, the presence of other \ion{C}{1} and \ion{Mg}{2} lines in the spectra suggests the possibility of line blending for the 1~\um\ feature. 
These complications motivated us to investigate the 2\um\ region.

%2mic results
2) Our sample of SESNe show two distinct groups based on their NIR properties.
These groups are dubbed NIR ``He-rich'' and ``He-poor'' and correspond almost perfectly to the optical SN~IIb/Ib and SN~Ic/Ic-BL groups, respectively.
Both the pEW and the PCA measurements show a dichotomy between the He-poor and He-rich groups (Figures \ref{fig:pEW} and \ref{fig:Ibc_PCA_proj}).
The dichotomy demonstrates that the 2~\um\ region is the ideal place to detect the presence of He.

% 2~\um\ cont.
3) Our investigations on the 2~\um\ region suggest that there are some trace amounts of He detected at early times in some He-poor SNe, confirming the results from previous works \citep[e.g.][]{Hachinger_2012, Piro_2014, Drout_2016, Prentice_2018}.  
Even though the \ion{He}{1} \lam2.0581~\um\ absorption in the He-poor SNe is much weaker compared to the He-rich, approximately half of the He-poor SNe in our sample show weak absorption features from this line.
Thus, there is perhaps some \ion{He}{1} contribution to the 1~\um\ feature in these SNe as well.

%implication of mass-loss
There are many ways that \SE\ can lose their outer envelopes including eruptions, radiation driven winds, stripping from common envelope in a binary system, fast rotations in Be stars, or a combination of them. 
One may expect that if the mass loss is from a single process, it would manifest as a continuous range of observed properties, indicating the degree of envelope stripping. 
NIR spectra provide a unique route to observe this, especially given the differences in the residual He. 
In this work, we find an abrupt difference in the 2~\um\ region between He-poor and He-rich SNe in our spectral and PCA analyses. 
This may indicate distinct mass-loss mechanisms for He-poor and He-rich SNe.
It may also be possible for a single mechanism to produce very different observational properties.
Further theoretical studies are therefore needed.
For instance, the He shell and core mass may be estimated by comparing NIR observations and models. 
Coupling this with stellar evolution information can place further constraints on the progenitor systems and the explosion mechanisms.

%%%%%%%%%%%%%%%%%%%%%%%%%%%%
\acknowledgements

We are pleased to acknowledge the following
individuals (in alphabetical order) for their assistance in obtaining
the NIR spectroscopic data set: Y.~Beletsky, T.~Dupuy, R.~Foley, L.~W.~Hsiao. 
We also thank the Las Campanas Observatory technical staff for their continued support over the years.
%CSP grants
The CSP-II has been supported by
NSF grants AST-1008343, AST-1613426, AST-1613455, and AST-1613472. 
%co-authors grants and ack. in alphabetical order
C.A. is supported by NASA grant 80NSSC19K1717 and NSF grants AST-1920392 and AST-1911074.
J.T. is funded by the consolidated STFC grant no. R276106.
L.G. acknowledges financial support from the Spanish Ministry of Science, Innovation and Universities (MICIU) under the 2019 Ram\'on y Cajal program RYC2019-027683 and from the Spanish MICIU project PID2020-115253GA-I00.
C.G. is supported by a Young Investor Grant (25501) from the VILLUM FONDEN.
Time domain research by D.J.S.\ is also supported by NSF grants AST-1821987, 1813466, 1908972, \& 2108032, and by the Heising-Simons Foundation under grant \#2020-1864. 
M.D.S. is supported by grants from the VILLUM FONDEN (grant number 28021) and  the Independent Research Fund Denmark (IRFD; 8021-00170B).

\software{\texttt{firehose} \citep{Simcoe_2013}, \texttt{SNID} \citep{Blondin&Tonry_2007}, \texttt{SNooPy} \citep{Burns_2011}, \texttt{xtellcor} \citep{Vacca_2003}}

%%%%%%%%%%%%%%%%%%Appendix
\appendix
% \restartappendixnumbering 
\renewcommand{\thefigure}{A\arabic{figure}}
\renewcommand{\theHfigure}{A\arabic{figure}}
\setcounter{figure}{0}
\renewcommand{\thetable}{A\arabic{table}}
\renewcommand{\theHtable}{A\arabic{table}}
\setcounter{table}{0}
\section{Classification} \label{appendix:Classification}

The optical classifications used in this work are listed in Table~\ref{tab:classification} in comparison to the previously published ones. 
The sources of optical spectra used are also listed. 
For each SN, the spectrum closest to maximum light was selected if it has adequate S/N ratio for classification.
The package SNID with the updated templates from \cite{Liu_Modjaz_2014} was then used for the classification.
Only slight differences were found comparing the classifications of this work with the previously reported ones. 
There were no SNe that changed between the IIb/Ib and Ic/Ic-BL groups.
Table~\ref{t:nir_class} lists all the SESNe in the sample that were classified using FIRE. 
Note that in the NIR, it is difficult to distinguish between SNe IIb and Ib.

%%%%%%%%%%
\begin{table}[h!]
    \caption{Optical spectroscopic classification}\label{tab:classification}
    \centering
    \begin{tabular}{ccccc}
    \hline
    \hline
    \multicolumn{1}{c}{\multirow{2}{*}{\bfseries SN}} &
    \multicolumn{2}{c}{\bfseries Optical Classification} & \multicolumn{2}{c}{\makebox[2pt]{\bfseries Source of Optical Spectrum}}\\
    &Published& This work$^{a}$& Published & This work\\\hline
ASASSN-14az			& IIb 	& IIb & \cite{ATEL6185}& \cite{Shivvers_2019} \\
LSQ13abf				& Ib & Ib& \cite{Stritzinger_2020}& CSP \\
LSQ13cum				& Ib & Ib& \cite{ATEL5561} & CSP	\\
LSQ13ddu				& Ibn & Ibn& \cite{Clark_2020} & CSP	\\
LSQ13doo 	& Ic-BL & Ic/Ic-BL & \cite{ATEL5615} & PESSTO 		\\
LSQ13lo					&Ib/c& Ic& \cite{2013ATel.4916....1H} & CSP	 \\
LSQ14akx				& IIb & IIb & \cite{ATEL5998}& CSP		 \\
OGLE-2014-SN-067 		& Ic & Ic-BL  &\cite{ATEL6430}& \cite{Childress_2016}	\\
iPTF13bvn				& Ib& Ib &\cite{Fremling_2016}& \cite{Cao_2013}	 \\
SN 2011hs				& IIb & IIb &\cite{Bufano_2014} &	CSP	\\
SN 2012J				& Ib/c & Ic & \cite{CBET2986}& CSP	\\
SN 2012P				& IIb & IIb & \cite{Fremling_2016}		& \cite{Fremling_2016}\\
SN 2012ap				& Ic-BL& Ic-BL &\cite{Milisavljevic_2015}  &	CSP	  \\
SN 2012au				& Ib & Ib &\cite{Milisavljevic_2013}&	\cite{Milisavljevic_2013}	\\
SN 2012hf				& Ic & Ic  & \cite{CBET3326} & PESSTO\\
SN 2012hn				& Ic & Ic  & \cite{CBET3337} & PESSTO\\
SN 2013ak				& IIb & IIb & \cite{2013ATel.4943....1M} & CSP	\\
SN 2013co				& Ic-BL & Ic & \cite{CBET3527}	&\cite{CBET3527}\\
SN 2013dk				& Ic & Ic 	& \cite{CBET3565}& \cite{ Elias-Rosa_2013}\\
SN 2013ek				& Ib & Ib &\cite{ATEL5227} &  PESSTO	\\
SN 2013el				& Ib-pec & IIb &\cite{ATEL5206} &	\cite{Shivvers_2019}	\\
SN 2013fq				& Ib & IIb & \cite{2013ATel.5400....1H}&	\cite{Childress_2016}	\\
SN 2013ge				& Ib/c & Ic & \cite{Drout_2016}	& \cite{Drout_2016}\\
SN 2014L				& Ic & Ic  & \cite{Zhang_2018} & CSP 		\\
SN 2014ad				& Ic-BL & Ic-BL	& \cite{Stevance_2017}& PESSTO	 \\
SN 2014ar				& Ic & Ic &\cite{ATEL6082} & CSP  		\\
SN 2014cp				& Ic-BL & Ic-BL & \cite{ATEL6302}& \cite{Childress_2016}	\\
SN 2014df				& Ib & Ib & \cite{ATEL6201}& \cite{Childress_2016}\\
SN 2014eh 	& Ic & Ic  &\cite{ATEL6667}& PESSTO	 \\
SN 2015Y		& IIb & Ib &\cite{ATEL7368} & \cite{Shivvers_2019}\\
    \end{tabular}
\end{table}

\begin{deluxetable}{cc}
\tablecolumns{3}
\tablecaption{FIRE NIR spectroscopic classifications of \SE.\label{t:nir_class}}
\tabletypesize{\small}
\tablehead{
\colhead{SN} &
\colhead{Source of classification}
}
\startdata
LSQ12fwb &\citet{2012ATel.4563....1H} \\
LSQ13lo&\citet{2013ATel.4916....1H} \\
OGLE-2013-SN-134&\citet{2013ATel.5664....1H} \\
OGLE-2014-SN-014&\citet{2014ATel.5891....1M} \\
PSN J11220840-3804001 &\citet{2014ATel.6220....1M} \\
SN~2013ak &\citet{2013ATel.4943....1M} \\
SN~2013dk&\citet{2013ATel.5167....1M} \\
SN~2013fq&\citet{2013ATel.5400....1H} \\
SN~2014df&\citet{2014ATel.6220....1M, 2014ATel.6442....1M} \\
\enddata
\end{deluxetable}

\renewcommand{\thefigure}{B\arabic{figure}}
\renewcommand{\theHfigure}{B\arabic{figure}}
\setcounter{figure}{0}
\renewcommand{\thetable}{B\arabic{table}}
\renewcommand{\theHtable}{B\arabic{table}}
\setcounter{table}{0}
\section{Spectra excluded from analyses}
\label{appendix:exclude}
\begin{figure}
\epsscale{1.22}
\centering
\plotone{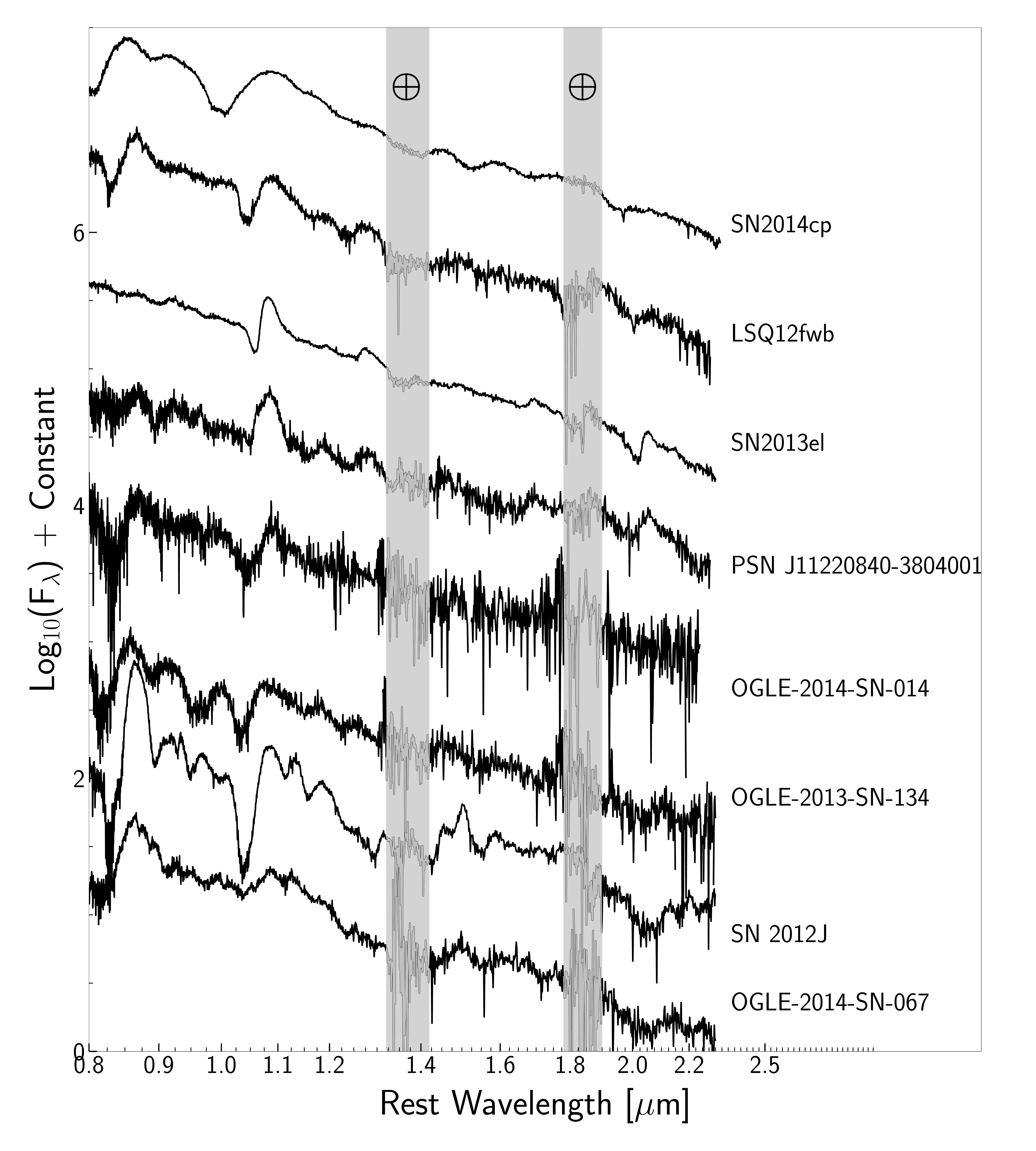}
\caption{Spectra of SNe with only one spectrum within the sample that do not have optical spectra and reliable photometry available.\label{fig:all_spec+bad}}
\end{figure}

In this section, we list the 8 NIR spectra of 8 SESNe excluded from our analyses, either because they do not have reliable phase estimates or optical spectroscopic classifications (Table~\ref{tab:1specs}). These SNe are presented in Figure~\ref{fig:all_spec+bad}.

\startlongtable
\begin{deluxetable}{ccccccc}
\tablecaption{SNe with no reliable phase or optical spectroscopic classification\label{tab:1specs}}
\tablehead{
\colhead{SN\tablenotemark{a}} & \colhead{Host} & \colhead{Redshift} &\colhead{UT date} & \colhead{MJD$_{obs}$} &
\colhead{Telluric STD} & \colhead{Airmass} }
\startdata
LSQ12fwb				&	CGCG 409-012      					&	0.0298& 2012-11-05 & 56236.6 & HD8325 		& 1.26\\
OGLE-2013-SN-134		&	WISEA J061355.87-725821.1				&	0.0382& 2013-12-14 & 56640.7 & HD19839 	& 1.39\\
OGLE-2014-SN-014		&	WISEA J042722.73-744206.6			&	0.0430\tablenotemark{b}& 2014-02-15 & 56703.5 & HD26493 	& 1.45\\
OGLE-2014-SN-067		&	anonymous							&	0.0187& 2014-09-03 & 56903.9 & HD42651 	& 1.48\\
PSN J11220840-3804001	&	ESO 319- G 016					&	0.0096& 2014-06-06 & 56814.6 & HD100852 	& 1.33\\
SN 2012J				&	ESO 386-IG 039					&	0.0094& 2012-02-03 & 55960.8 & HD125062 	& 1.52\\
SN 2013el				&	NGC 1285						&	0.0175& 2013-07-16 & 56489.9 & HD25266 	& 1.27\\
SN 2014cp				&	ESO 479- G 001					&	0.0162& 2014-07-10 & 56848.9 & HD16636 	& 1.22\\
\enddata
\tablenotetext{a}{Redshift obtained from \cite{2014ATel.5891....1M}}
\end{deluxetable}

\renewcommand{\thefigure}{C\arabic{figure}}
\renewcommand{\theHfigure}{C\arabic{figure}}
\setcounter{figure}{0}
\renewcommand{\thetable}{C\arabic{table}}
\renewcommand{\theHtable}{C\arabic{table}}
\setcounter{table}{0}

\begin{figure}
\epsscale{1.3}
\centering
\plotone{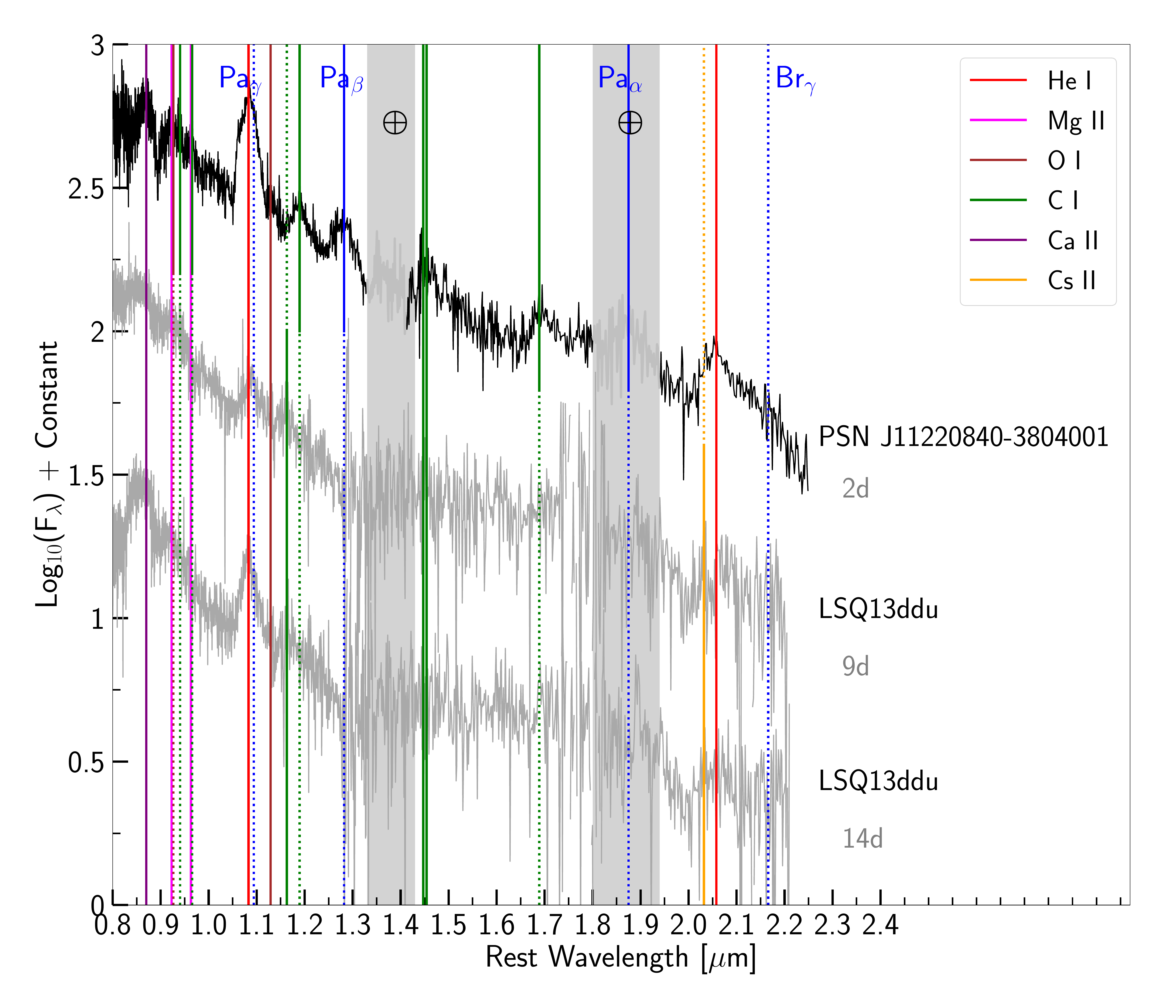}
\caption{Comparing spectra of PSN J11220840-3804001 with LSQ13ddu. PSN J11220840-3804001 is plotted in black and LSQ13ddu with 2 spectra is plotted in grey.\label{fig:csp14abs}}
\end{figure}

\section{Notable objects}

\paragraph{SN~2014df}
SN~2014df has an unusual feature at $\sim$2~\um, shown in Figure \ref{fig:paperplots_4X4}, that makes investigating this SN interesting.
This feature at $\sim$2~\um\ shows two components in the earliest spectrum. 
This two component feature reverts to a single apparent component in the subsequent spectrum, taken 52 days afterwards. 
SN~2014df is the only SN in this sample that shows this unusual double component feature. 
In order to identify this feature, we measured the \ion{He}{1} velocity of the blue component of this feature to be at $14{,}400$~\kms\ and the red component to be $8{,}200$~\kms. 
Assuming \ion{Mg}{2} the blue component gives a velocity of $24{,}900$~\kms, and the red component gives $19{,}000$~\kms. 
Since we observe a really high velocity by assuming \ion{Mg}{2}, it is unlikely that either of these two components are formed by \ion{Mg}{2}. 
Therefore, both components could possibly be a result of \ion{He}{1} with one of the components formed by high velocity \ion{He}{1}.

\paragraph{PSN J11220840-3804001}
It is worth to note that, there is one SN in the sample, PSN J11220840-3804001, that has a very similar spectrum to the NIR spectra of LSQ13ddu. 
\cite{Clark_2020} classified LSQ13ddu as a SN~Ibn.
Figure~\ref{fig:csp14abs} compares the two SNe with some of the features identified.

\section{Gaussian fits}\label{Appendix:fits}

\begin{figure*}
\epsscale{1}
\centering
\plotone{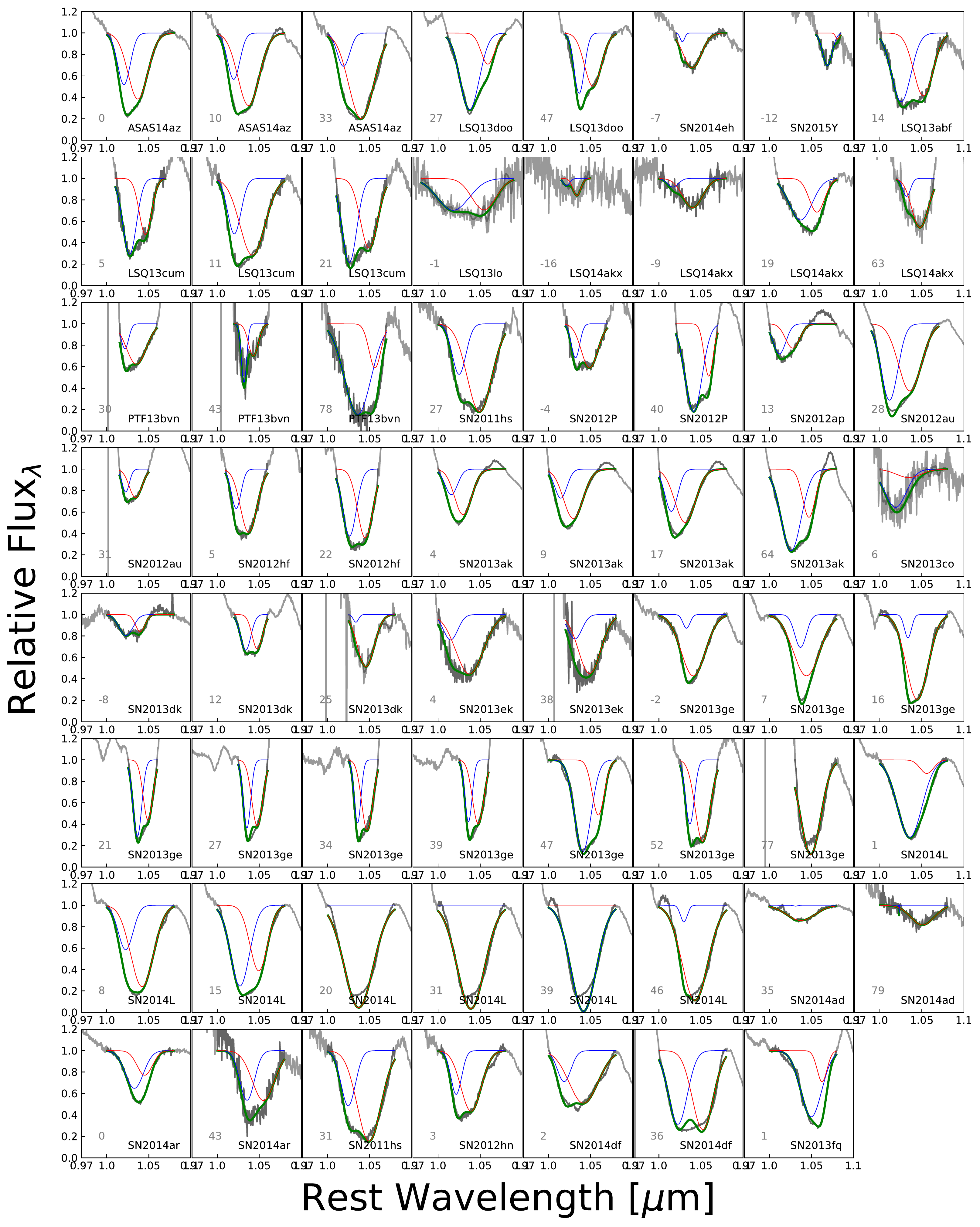}
\caption{The two-Gaussian fits for the 1~\um\ region of all the spectra used in the analysis. The two earliest spectra of SN~2014ad were excluded due to the fact that there is not a strong enough feature in these two spectra to fit. LSQ13ddu was excluded due to low S/N. For cases that a minimum was not found for one of the components, the minimum of the entire feature was considered for that component instead (Figure \ref{fig:velocities_comps}). \label{fig:1fits}}
\end{figure*}

\begin{figure*}
\epsscale{1}
\centering
\plotone{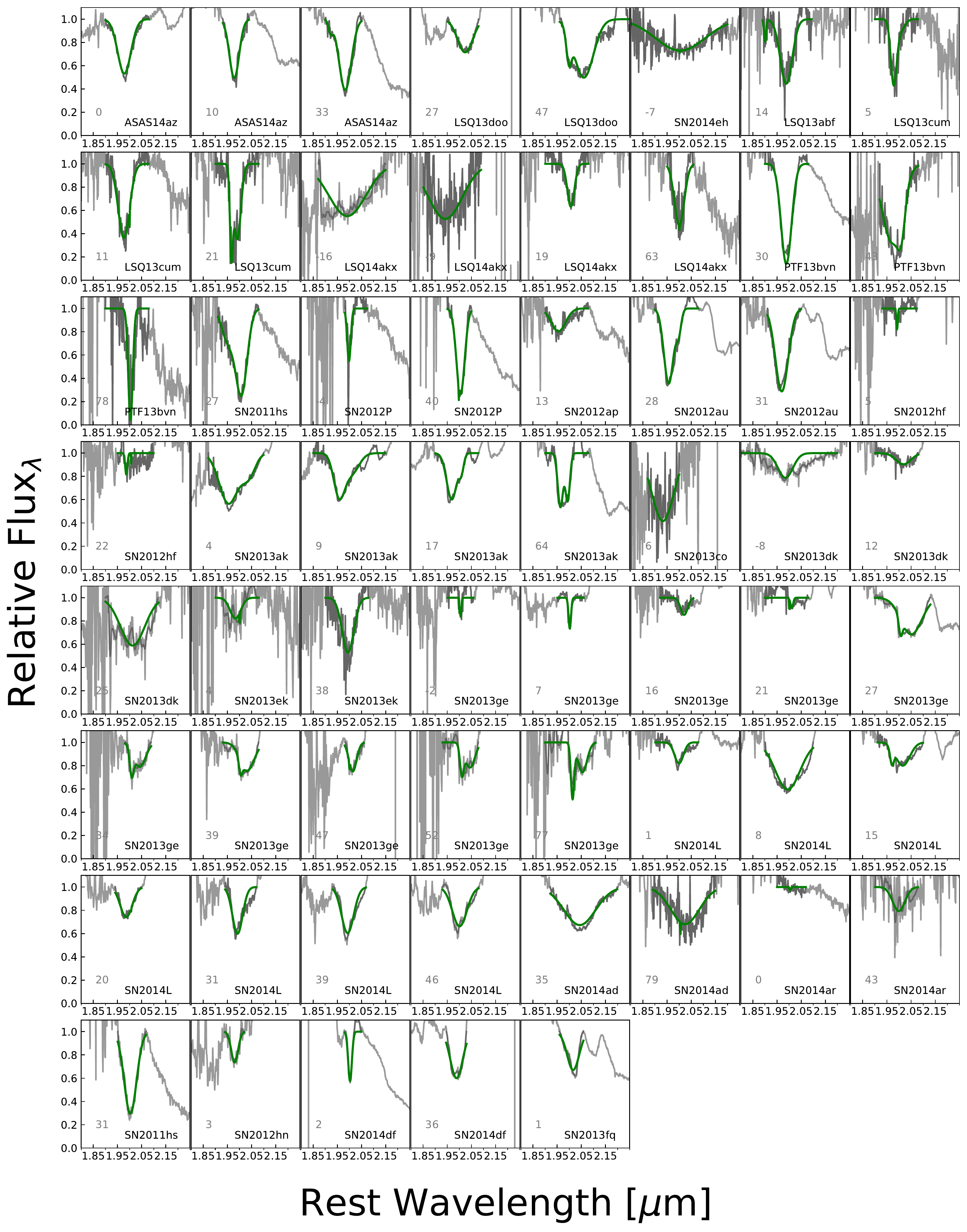}
\caption{The two-Gaussian fits for the 2~\um\ region of all the spectra used in the analysis. The two earliest spectra of SN~2014ad was excluded due to the fact that there is not a strong enough feature in these spectra to fit. The SNe with S/N lower than 5 in the 2~\um\ region were also excluded. Since we did not measure and use the minimum of each component individually for this feature, the two Gaussian components are not shown.\label{fig:2fits}}
\end{figure*}

Figure~\ref{fig:1fits} and Figure~\ref{fig:2fits} present the Gaussian fits done to measure velocities.

\end{document}